\documentclass[apj]{emulateapj}
\usepackage{natbib}

\usepackage{color}

\newcommand{\simgt}{\lower.5ex\hbox{$\; \buildrel > \over \sim \;$}}
\newcommand{\simlt}{\lower.5ex\hbox{$\; \buildrel < \over \sim \;$}}

\newcommand{\Yx}{Y_{\rm X}}
\newcommand{\Mgas}{M_{\rm gas}}
\def\ls{\mathrel{\hbox{\rlap{\hbox{\lower4pt\hbox{$\sim$}}}\hbox{$<$}}}}
\def\gs{\mathrel{\hbox{\rlap{\hbox{\lower4pt\hbox{$\sim$}}}\hbox{$>$}}}}

\shorttitle{LoCuSS: Calibrating Mass-Observable Scaling Relations for Cluster Cosmology}
\shortauthors{Okabe, Zhang, Finoguenov et al.}

\journalinfo{ApJ}
\submitted{ApJ, accepted}

\begin{document}

\title{LoCuSS: Calibrating Mass-Observable Scaling Relations for Cluster Cosmology
  with Subaru Weak Lensing Observations \altaffilmark{*}}
\altaffiltext{*}{This work is based in part on data collected at
  the Subaru Telescope and obtained from the SMOKA, which is operated by
  the Astronomy Data Center, National Astronomical Observatory of
  Japan.  Based on observations made with the \emph{XMM-Newton}, an
  ESA science mission with instruments and contributions directly
  funded by ESA member states and the USA (NASA).}


\author{
  N.\ Okabe,$\!$\altaffilmark{1,2}
  Y.-Y.\ Zhang,$\!$\altaffilmark{3}
  A.\ Finoguenov,$\!$\altaffilmark{4,5}
  M.\ Takada,$\!$\altaffilmark{6}
  G.\ P.\ Smith,$\!$\altaffilmark{7}
  K.\ Umetsu,$\!$\altaffilmark{2}
  T.\ Futamase,$\!$\altaffilmark{1}
}

\email{okabe@asiaa.sinica.edu.tw}
\altaffiltext{1}{Astronomical institute, Tohoku University, Aramaki,
  Aoba-ku, Sendai, 980-8578, Japan} 
\altaffiltext{2}{Academia Sinica Institute of Astronomy and
  Astrophysics (ASIAA), P.O. Box 23-141, Taipei 10617, Taiwan}
\altaffiltext{3}{Argelander-Institut f\"ur Astronomie, Universit\"at
  Bonn, Auf dem H\"ugel 71, 53121 Bonn, Germany}
\altaffiltext{4}{Max-Planck-Institut f\"ur extraterrestrische Physik,
  Giessenbachstra\ss e, 85748 Garching, Germany}
\altaffiltext{5}{University of Maryland, Baltimore County, 1000
  Hilltop Circle, Baltimore, MD 21250, USA} \altaffiltext{6}{Institute
  for the Physics and Mathematics of the Universe (IPMU), The
  University of Tokyo \\ 5-1-5 Kashiwa-no-Ha, Kashiwa City, Chiba
  277-8568, Japan} \altaffiltext{7}{ School of Physics and Astronomy,
  University of Birmingham, Edgbaston, Birmingham, B15 2TT, UK}

\begin{abstract}
  We present a joint weak-lensing/X-ray study of galaxy cluster
  mass-observable scaling relations, motivated by the critical
  importance of accurate calibration of mass proxies for future X-ray
  missions, including \emph{eROSITA}. We use a sample of 12 clusters
  at $z\simeq0.2$ that we have observed with Subaru and
  \emph{XMM-Newton} to construct relationships between the
  weak-lensing mass ($M$) and three X-ray observables: gas
  temperature ($T$), gas mass ($\Mgas$), and quasi-integrated gas
  pressure ($\Yx$) at overdensities of $\Delta=2500$, $1000$, and
  $500$ with respect to the critical density.  We find that $\Mgas$ at
  $\Delta\le1000$ appears to be the most promising mass proxy of the
  three because it has the lowest intrinsic scatter in mass at fixed
  observable, $\sigma_{\ln\!M}\simeq0.1$, independent of the cluster
  dynamical state.  The scatter in mass at fixed $T$ and $\Yx$ is a
  factor of $\sim2-3$ larger than at fixed $\Mgas$, which are
  indicative of the structural segregation that we find in the $M-T$
  and $M-\Yx$ relationships.  Undisturbed clusters are found to be
  $\sim40\%$ and $\sim20\%$ more massive than disturbed clusters at
  fixed $T$ and $\Yx$ respectively at $\sim2\sigma$ significance. 
In particular, A\,1914 -- a well-known merging cluster -- significantly
increases the scatter and lowers the normalization of the relation
for disturbed clusters.  We also investigated the covariance between
intrinsic scatter in $M-\Mgas$ and $M-T$ relations, finding that  
they are \emph{positively} correlated.
 This contradicts the
  adaptive mesh refinement simulations that motivated the idea that
  $\Yx$ may be a low scatter mass proxy, and agrees with more recent
  smoothed particle hydrodynamic simulations based on the Millennium
  Simulation.
We also propose a method to identify a robust mass proxy based on  
principal component analysis.
  The statistical precision of our results is limited by
  the small sample size and the presence of 
 the extreme merging cluster
  in our sample.  We therefore look forward to studying a larger, more
  complete sample in the future.
\end{abstract}

\keywords{Cosmology: observations -- dark matter -- galaxies: clusters: general --
  gravitational lensing: weak -- X-rays: galaxies: clusters.
  }

\section{Introduction}


Galaxy clusters are the largest virialized objects in the universe;
they formed from high amplitude peaks of the primordial density
field.  Clusters therefore occupy the high mass exponential tail of
the dark matter halo mass function, which is sensitive to the matter
density and expansion history of the universe, and to modifications of
the laws of gravity.  Measurements of the evolution of the galaxy cluster
mass function across a broad range of redshifts can thus provide a
powerful tool for constraining the cosmological parameters (e.g.,
Vikhlinin et al. 2009a, 2009b).  Numerous galaxy cluster surveys will
soon begin delivering a huge amount of data at optical, X-ray, and
millimeter wavelengths, e.g.\ from Subaru/Hyper-Suprime-Cam,
\emph{eROSITA}, \emph{SPT} and \emph{ACT}.  One of the main goals of
these surveys is to measure the evolution of the galaxy cluster mass
function, and thus to probe the expansion history of the universe.
However, the mass of a galaxy cluster is not directly measurable.
These surveys will therefore rely on ``mass-like'' observables (e.g.,\
X-ray temperature -- Evrard et al.\ 1996) and scaling relations
between these observables and mass, to construct the all-important
mass functions.  Calibration of mass-observable scaling relations is
therefore currently a high priority observational goal.

Traditionally, observational studies of the mass-observable scaling
relations have relied solely on X-ray observations, typically
concentrating on the mass-temperature relation (e.g.,\ Finoguenov et
al., 2001; Sanderson et al., 2003; Ettori et al., 2004; Arnaud et al.,
2005).  X-ray-based mass measurements require hydrostatic equilibrium
(H.E.) and spherical symmetry to be assumed, and either measurement of
the temperature profile, or an assumption of isothermality.
Inclusion of X-ray temperature information in both axes of the
mass-temperature relation may therefore induce intrinsic correlations
into the measured relation.  The validity of the underlying
assumptions also warrants careful testing.

Gravitational lensing offers cluster mass measurements that are
independent of X-ray observations, and do not rely on
assuming H.E..  Joint lensing/X-ray studies (e.g., Okabe \& Umetsu 2008; Kawaharada et
al. 2010) are therefore a promising
route for calibrating cluster mass-observable scaling relations.
Indeed, early lensing/X-ray studies of cluster cores indicated that
the scatter in cluster temperature may be as large as $40\%$ at fixed
mass, and that the scatter is dominated by disturbed, merging
clusters, in which H.E.\ may not hold (Smith et al.\ 2005).
Subsequent work has concentrated on using weak-lensing data to extend
this pioneering work beyond cluster cores to overdensities of
$500\ls\Delta\ls2500$ with respect to the critical density (Bardeau et
al.\ 2007; Hoekstra 2007; Pedersen \& Dahle 2007;
Zhang et al.\ 2007, 2008).  The main limiting factors in these weak-lensing/X-ray studies
have been the limited statistical precision and heterogeneity of the
available weak-lensing data, and also the small samples observed to
date.

On the theoretical side, Kravtsov et al.\ (2006) proposed the
so-called quasi-integrated pressure, $\Yx{\equiv}M_{\rm
  gas}{\times}T$ as a ``new robust low-scatter X-ray mass indicator'',
or, a mass-like observable.  This was motivated by analysis of their
hydrodynamic numerical simulations of clusters using an adaptive mesh
refinement (AMR) code.  They found that the temperature deviations
from the $M-T$ relation are anti-correlated with the gas mass
deviations from the $M-\Mgas$ relation.  This anti-correlation found
in their simulations acts to suppress the scatter in the $M-\Yx$
relation, independent of the dynamical state of the clusters.  This
prediction has stimulated much observational effort within the X-ray
community that has broadly supported the idea that $\Yx$ is the
optimal X-ray mass proxy (e.g.,\ Maughan 2007; Arnaud et al., 2007;
Vikhlinin et al.\ 2009a).

However, Stanek et al.'s (2010) smoothed particle hydrodynamic (SPH)
Millennium Gas Simulations contradict Kravtsov et al.'s simulations.
Stanek et al.\ predict that the temperature and gas-mass deviations
are positively correlated; this result appears to be independent of
the range of gas physics (gravity only, cooling, preheating)
implemented in the simulations.  Juett et al.\ (2010) have also
recently suggested that previous X-ray-only studies may have
underestimated the scatter in mass-observable scaling relations by a
factor of $\sim2-3$.  In summary, a joint lensing/X-ray observational
investigation of the relationships between mass and gas mass,
temperature, and $\Yx$, is urgently needed.  Such joint studies also
lend themselves well to the task of observationally testing various
corrections that have been derived from numerical simulations to
account for deviations from H.E..  For example, numerous authors have
pointed out that H.E.\ mass estimates may underestimate the cluster
mass because of non-thermal pressure support due to turbulence caused
by bulk motion of the cluster gas (e.g., Evrard 1990; Rasia et al.\
2006; Nagai et al.\ 2007; Piffaretti \& Valdarnini 2008; Fang et al.\
2009), and Vikhlinin et al.\ (2009a) applied a $17$\% upward
correction to X-ray masses of disturbed clusters, based on the results
of simulations.

A key goal of the Local Cluster Substructure Survey
(LoCuSS\footnote{\sf http://www.sr.bham.ac.uk/locuss}) is to calibrate
cluster mass-observable scaling relations for future cosmological
experiments.  LoCuSS is a multi-wavelength survey of galaxy clusters
at $0.15<z<0.3$ selected from the \emph{ROSAT} All-sky Survey catalogs
(Ebeling et al.\ 1998, 2000; B\"ohringer et al.\ 2004).  To date we
have published the first lensing/Sunyaev-Zeldovich effect comparison
(Marrone et al.\ 2009), begun our lensing/X-ray scaling relation work
with a pilot study (Zhang et al.\ 2008), and compared lensing-based
masses with H.E.\ masses on both small (Richard et al., 2010) and
large (Zhang et al.\ 2010) scales.  This article is a continuation of
our pilot study (Zhang et al.\ 2008), in which we combined
weak-lensing mass measurements from the Canada-France-Hawaii Telescope (Bardeau et al.\ 2005, 2007)
and from the Nordic Optical Telescope and UH 88in (Dahle 2006) with
\emph{XMM-Newton} observations to calibrate the mass-observable
scaling relations.  As alluded to above, Zhang et al.'s results were
limited by the quality of the weak-lensing mass measurements, because
the underlying data were heterogeneous in observing facilities, fields
of view, and filters used.  In this article we address these issues by
using our own weak-lensing mass measurements based on uniform analysis
of our Subaru/Suprime-cam observations (Okabe \& Umetsu 2008; Okabe et
al., 2010).  Nevertheless, our Subaru/\emph{XMM-Newton} sample remains
small, at just 12 clusters.  As we discuss throughout this article,
sample size therefore remains an issue, and we will address this in a
future article.

The outline of this paper is as follows. In Sec.~\ref{sec:analysis} we
briefly describe the weak lensing and X-ray analysis, and measure the
dynamical state of each cluster using \emph{XMM-Newton} data.  We
present the main results on the mass-observable scaling relations in
Sec.~\ref{sec:results}, discuss the results in Sec.~\ref{sec:discuss},
and summarize our work in Sec.~\ref{sec:sum}.  Throughout this paper,
we assume $\Omega_{m,0}=0.3$, $\Omega_\Lambda=0.7$, and $h=H_0/100~{\rm kms^{-1}Mpc^{-1}}=0.7$.

\section{Sample and data analysis} \label{sec:analysis}

\subsection{Sample}


For the purpose of this paper, we compiled a sample of 12 clusters --
A\,68, A\,115, A\,209, A\,267, A\,383, A\,1835, A\,1914, Z\,7160,
A\,2261, RX\,J2129.6$+$0005, A\,2390, and A\,2631 -- that represents
the overlap between the samples for which Subaru/Suprime-Cam and
\emph{XMM-Newton} data are available, and that we have previously
published (Zhang et al.\ 2008; Okabe \& Umetsu 2008; Okabe et al.\
2010).  The sample does not suffer, by design, any strong biases to
extreme merging or extreme cool core clusters, and therefore can be
regarded, qualitatively, as representative of massive, X-ray luminous
clusters.  However, given the small sample size, we refrain from
attempting to quantify how these 12 might be biased with respect
to the underlying cluster population in this article.  Instead, this
article presents some early results from our Subaru/\emph{XMM-Newton}
program, that benefit from the use of our Subaru data, as opposed to
the CFH12k/UH8k/NOT data that we used in Zhang et al.\ (2008).  We
defer detailed discussion of sample definition and possible biases to
future articles in this series that will address larger, more complete
samples.

\subsection{Weak-lensing mass measurements}


The details of our weak-lensing analysis are described in by Okabe \&
Umetsu (2008), and Okabe et al.\ (2010); here we provide a brief
outline of some important aspects of our methods.



We selected background galaxies based on their location in the
color-magnitude plane -- typically $(V-i^\prime)/i^\prime$ -- bluer or
redder than cluster red-sequence by a minimum color-offset (Umetsu \&
Broadhurst 2008; Umetsu et al.\ 2009; Okabe et al.\ 2010).  As
demonstrated by Okabe et al.\ (2010), contamination of the background
galaxy catalogs by faint (unlensed) cluster members dilutes the
weak-lensing signal.  This effect is more pronounced at smaller
clustercentric radii because the number density of cluster galaxies
rises towards
 the cluster centers.  In the absence of our
color-selection techniques, weak-lensing $M_{500}$ and $M_{2500}$
measurements can be biased low by $\sim20$\%--$50$\%.

We used the COSMOS photometric redshift catalog (Ilbert et al.\ 2009)
to estimate the redshift of the background galaxies.  Specifically, we
calculated the average lensing weight, $\langle D_{\rm LS}/D_{\rm
  OS}\rangle=\int_{z_d}dz d P_{\rm WL}/dz D_{\rm LS}/D_{\rm OS}$ (see
also Equation (10) in Okabe et al. 2010), of each background galaxy 
catalog
by
selecting galaxies identical to both our catalogs, and the COSMOS
catalog.  $D_{\rm OS}$ and $D_{\rm LS}$ are the angular diameter
distances between the observer and source (background galaxy) and lens
and source respectively.

In cosmology the three-dimensional spherical mass, $M_\Delta$, enclosed within a
sphere of radius $r_\Delta$ for a given overdensity $\Delta$ is most
relevant for the cluster mass function, where $r_\Delta$ is chosen
such that the average density within the sphere is equal to the
critical mass density at the cluster redshift, $\rho_{\rm cr}$, times
the overdensity $\Delta$.  We estimated $M_\Delta$ for each cluster by
fitting the measured radial profile of lensing distortion signals to
the NFW model prediction parameterized by the mass $M_\Delta$ and
$c_\Delta$, where the NFW mass profile (Navarro, Frenk \& White 1996,
1997) is given as $\rho \propto r^{-1}(1+c_\Delta r/r_{\Delta})^{-2}$
with $c_\Delta$ being the concentration parameter.

Describing cluster-scale dark matter halos as spherical objects may
cause systematic errors in individual mass measurements because
clusters are predicted to be triaxial in the collisionless CDM model
(Jing \& Suto 2002).  For example, if the major axis of a triaxial
halo is aligned with or perpendicular to the line of sight, a
spherical model would overestimate or underestimate the mass, respectively,
and also cause systematic errors in the measurement of the
concentration parameter (Oguri et al.\ 2005; Gavazzi\ 2005;
Corless et al.\ 2009).  However, if the distribution of cluster
orientations is random, then adopting spherical mass models should not
introduce a significant bias into the properties of the sample.  We
therefore check that this is the case for our sample by comparing the
spherical mass measurements from Okabe et al.\ (2010) that we use here
with triaxial mass measurements of the same clusters using the same
background galaxy catalogs from Oguri et al.\ (2010).  On average the
spherical ($M_{\Delta}^{\rm(sph)}$) and triaxial ($M_{\Delta}^{\rm(tri)}$)
masses agree 
well -- $\langle M_{\Delta}^{{\rm(tri) }}/M_{\Delta}^{{\rm
    (ave)}}\rangle=0.98\pm0.15, 0.90\pm0.17~{\rm and}~0.83\pm0.21$, for
    $\Delta=500,1000~{\rm and}~2500$
 -- confirming the expectation of negligible bias.

\subsection{X-ray observables} \label{subsec:xray}



The observations and data reduction are described
in detail by Zhang et al.\ (2007, 2008).  
In brief, the three mass
proxies considered in this article are calculated as follows.  The
global temperature is a volume average of the spectrally measured,
radial temperature profile limited to the radial range of
$(0.2-0.5)r_{500}$.  The gas mass $\Mgas(r)$ was obtained for each
cluster by integrating a double-$\beta$ model of the electron density
that was fitted to the X-ray surface brightness profile.  The
quasi-integrated pressure is the product of the gas mass and the
global temperature: $\Yx(r) = M_{\rm gas}(r)\times
T_{0.2-0.5r_{500}}$.  Note that  $M_{\rm gas} (r)$,
$T_{0.2-0.5r_{500}}$ and $\Yx(r)$ have been calculated using radii
obtained from the weak-lensing analysis, and not using radii
calculated from the X-ray analysis as in Zhang et al.\ (2008).  This
definition of radii introduces a subtle correlation with weak lensing
mass -- we will explore this when estimating the intrinsic scatter in
the mass-observable scaling relations in Sec.\ \ref{subsec:cov_int} and
the Appendix.
Finally, we adopted a self-consistent definition of the cluster
centers based on the weak lensing analysis.  This caused us to change
the centers of just two clusters -- A\,1914 and A\,2631 -- from those
used by Zhang et al.\ (2010). 
   
\subsection{X-ray morphology and dynamical state} \label{sec:a+f}



Previous joint lensing/X-ray studies have identified the dynamical
state of clusters as a significant source of scatter in
mass-observable scaling relations (Smith et al.\ 2005; Pedersen \&
Dahle 2007; Zhang et al.\ 2008, 2009).  In this section we therefore
classify the clusters as either ``disturbed'' or ``undisturbed'',
based on a new method patterned on those developed for the
morphological classification of galaxies (e.g.,\ Conselice 2003).

We calculate the asymmetry ($A$) and fluctuation ($F$) of the X-ray
surface brightness distribution in the $0.7-2$~keV band.  Asymmetry is
defined as $A=(\sum_{ij}|I_{ij}-R_{ij}|)/\sum_{ij}I_{ij}$, the
normalized sum of the absolute value of the flux residuals where
$I_{ij}$ is a matrix element of the combined MOS1+MOS2
\emph{XMM-Newton} frame in the 0.7-2.0~keV band, flat fielded, point
source subtracted and refilled assuming a Poisson distribution, and
$R_{ij}$ are the matrix elements obtained by rotating the above frame
by $180^\circ$.  The pixel size of both frames is
$4^{\prime\prime}\times 4^{\prime\prime}$.  The fluctuation, $F$, measures
deviations from a smooth flux distribution and is defined as
$F=(\sum_{ij}I_{ij}-B_{ij})/\sum_{ij} I_{ij}$, where $B_{ij}$ is an 
element in a frame smoothed on 2\,arcmin scales, which corresponds to
a physical scale of $400$~kpc at $z=0.2$.  Such smoothing also
suppresses the effect of the complex shape of the \emph{XMM-Newton} point
spread function (Ghizzardi 2001).  We estimate the statistical
errors of $A$ and $F$ assuming Poisson noise computed within a radius
of $r_{500}$, excluding CCD gaps and bad pixels.  We also estimate the
systematic error of $A$ caused by uncertainties in the cluster centers
by recalculating $A$, each time moving the cluster centers onto one of
the neighboring pixels within the $r\le 4^{\prime \prime}$ circle from
the nominal cluster center. 

\begin{figure}
\begin{center}
\includegraphics[angle=270,width=7cm]{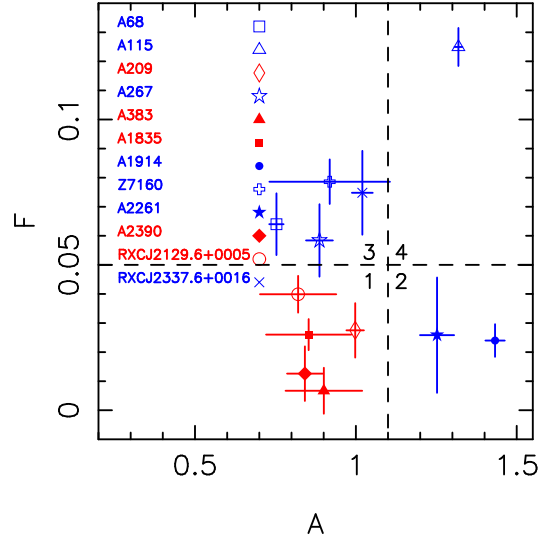}
\end{center}
\caption{Asymmetry versus fluctuation parameters using the $\le
  r_{500}$ region. We define undisturbed clusters (red, low A and F
  parameters) and disturbed clusters (blue, high A parameter or high F
  one). }
\label{f:a+f}
\end{figure}

The clusters span the range $A\sim0.07-0.15$ and $F\sim0-0.14$
(Fig.~\ref{f:a+f}).  Dynamically disturbed clusters generally have an
asymmetric X-ray morphology, with an offset between optical and X-ray
centers, and are therefore expected to have larger $A$ and $F$ than
undisturbed clusters.  To separate the clusters into two subsamples
that represent relatively disturbed and relatively undisturbed
systems, we subdivided the $A-F$ plane into four quadrants: (1) $A<1.1$
and $F<0.05$ -- RX\,J2129, A\,209, A\,383, A\,1835, and A\,2390, (2)
$A>1.1$ and $F<0.05$ -- A\,2261 and A\,1914, (3) $A<1.1$ and $F>0.05$
-- A\,68, A\,2631, A\,267, and Z\,7160, and (4) $A>1.1$ and $F>0.05$
-- A\,115. We classify the five clusters in
quadrant (1) -- low $A$ and low $F$ -- as undisturbed clusters, and
the remaining seven as disturbed clusters.  It is immediately obvious
that this classification matches other possible classification schemes
well.  For example, four of the five undisturbed clusters host a cool
core (e.g.\ Smith et al.\ 2003; Allen et al.\ 2001; Peterson et al.\
2002), and the disturbed clusters have been discussed extensively as
merging/cold-front clusters (e.g., Okabe \& Umetsu 2008; Mazzotta \&
Giacintucci 2008; Gutierrez \& Krawczynski 2005), in which complicated
temperature/entropy distributions or large offsets between
lensing/optical and X-ray centroids exits (e.g., Finoguenov et al.\ 2005;
Smith et al.\ 2005; Sanderson et al.\ 2009a).  In summary, all of the
clusters identified as disturbed in the $A-F$ plane are independently
confirmed as disturbed by other methods in the literature.  However we
stress again the relative nature of the disturbed/undisturbed
classification, and acknowledge that the disturbed clusters in
particular likely comprise clusters in a wide variety of stages in
their dynamical evolution.  We will return to this issue later when we
assess the impact of a single extreme merging cluster on our attempts
to calibrate the mass-observable scaling relations.


\section{Results}\label{sec:results}


In this section, we present the main empirical results of the slope,
normalization, and intrinsic scatter in the mass-observable scaling
relations and how these depend on the dynamical state of the
clusters.  We also discuss the correlation between gas mass and
temperature deviations.

\subsection{Scaling relations and fitting methods} \label{sec:m-o}

If gravitational heating is the dominant mechanism responsible for the
X-ray properties of galaxy clusters, the following scaling relations
are expected to hold:
\begin{eqnarray}
M E(z)~&\propto&~({\Yx}E(z))^{3/5}\,h^{1/2} , \label{eq:m-yx}\\
M E(z)~&\propto&~M_{{\rm gas}}E(z)\,h^{3/2} , \label{eq:m-mgas}  \\
M E(z)~&\propto&~T^{3/2}\,h^{-1} , \label{eq:m-t} 
\end{eqnarray}
where $M, \Mgas$, and $T$ are the total mass, gas mass, and
temperature of a cluster, respectively, and $\Yx=\Mgas \times T$ is the
quasi-integrated pressure.  These relations, specifically the
exponents of $M, \Mgas$, and $T$, are usually referred to as
self-similar, following Kaiser (1986).  Note that the term
$E(z)=H(z)/H_0=[\Omega_{m,0}(1+z)^3+\Omega_\Lambda]^{1/2}$ accounts
for the redshift evolution of the clusters in a flat universe.

In the following subsections we therefore fit the functional form
$M_z=M_0X_z^\gamma$ to the data, where $M_z=M\,E(z)$, $M_0$ is the
normalization, $X_z$ is the X-ray observable (i.e.,\ $\Yx$, $T$, or
$\Mgas$) multiplied by $E(z)$ or not as appropriate based on Equations
1-3, and $\gamma$ is the logarithmic slope.  These fits are done at
three overdensities with respect to the critical density:
$\Delta=2500, 1000$ and $500$. The scaling relation slope and
normalization measurements are based on orthogonal regression
performed using the Orthogonal Distance Regression package (ODRPACK,
e.g.\ Boggs et al.\ 1987) taking into account the measurement errors.
In general we ignore the subtle correlations introduced by measuring
the X-ray observables within radii defined by the weak-lensing
analysis, although we do take them into account in Sec.\ref{subsec:scatter} when we
measure the intrinsic scatter.  To check for consistency with other
work, we have also refitted the relations using the bisector
modification of the BCES method (Akritas \& Bershady 1996).  The
difference on best-fit scaling relation parameters between the two
fitting methods is a small fraction of statistical uncertainties.  For
example, the difference on best-fit slopes and normalizations between
the two methods is typically $\sim30\%$ and $\sim6\%$ of the
statistical error respectively.
 We also did the bootstrap resampling
 to estimate the
 sample variance on the slope parameter, and found that it is
 $\ls20\%$ of the statistical errors. 

\subsection{Slope and normalization} \label{subsec:norm}

\begin{figure*}
  \includegraphics[angle=270,width=5.5cm]{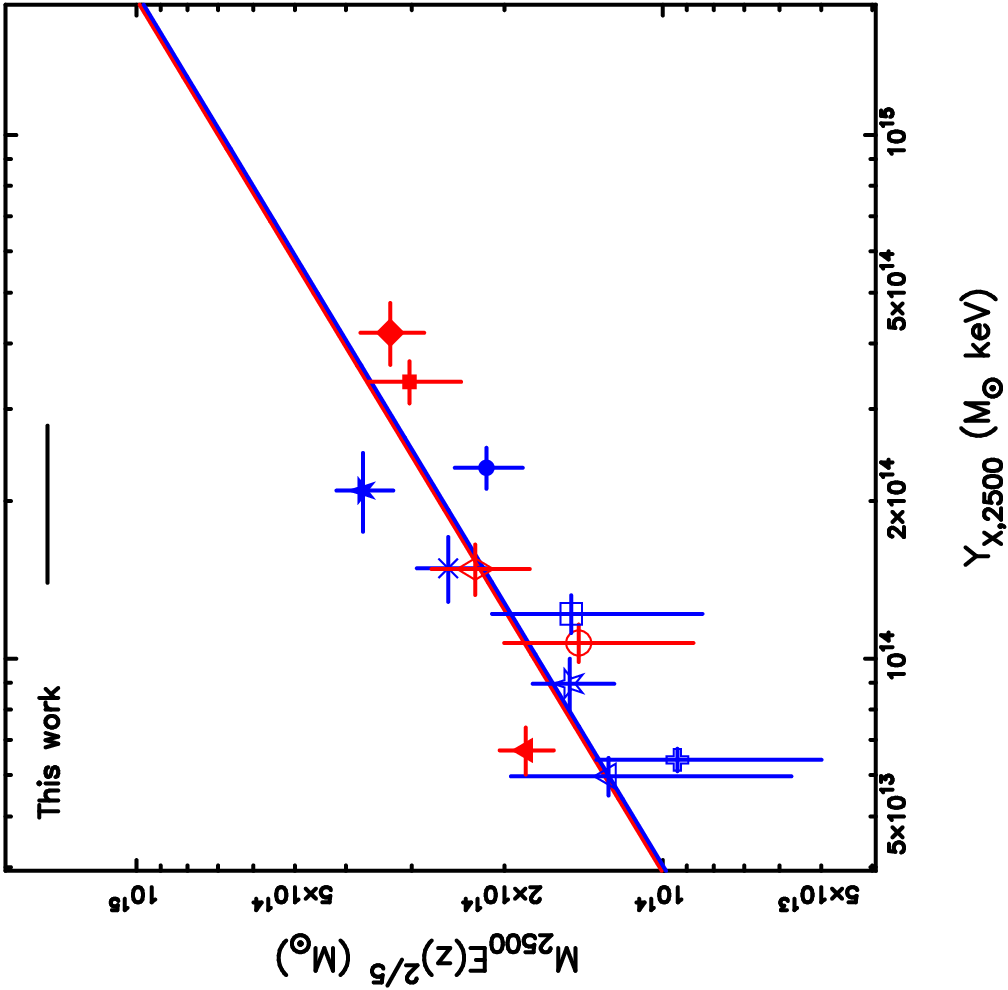}
  \hspace{-0.3cm}
  \includegraphics[angle=270,width=5.5cm]{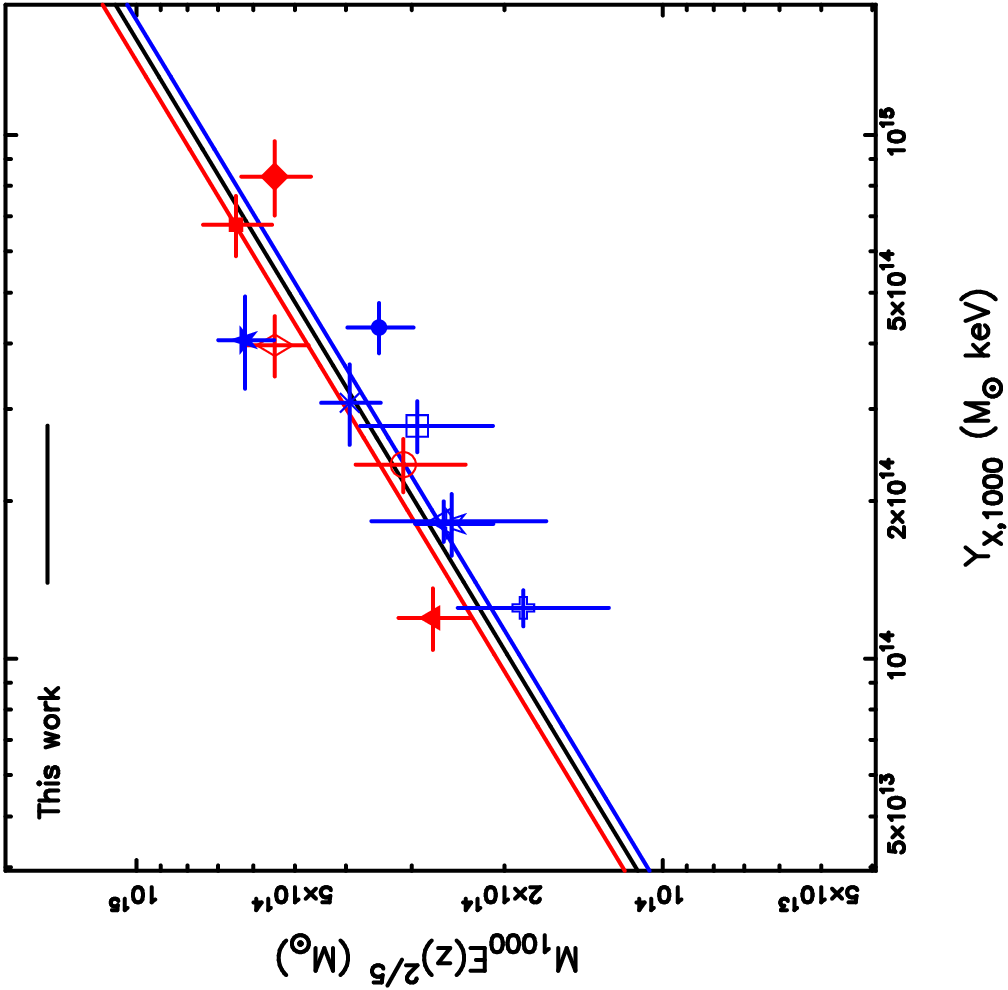}
  \hspace{-0.3cm}
  \includegraphics[angle=270,width=5.5cm]{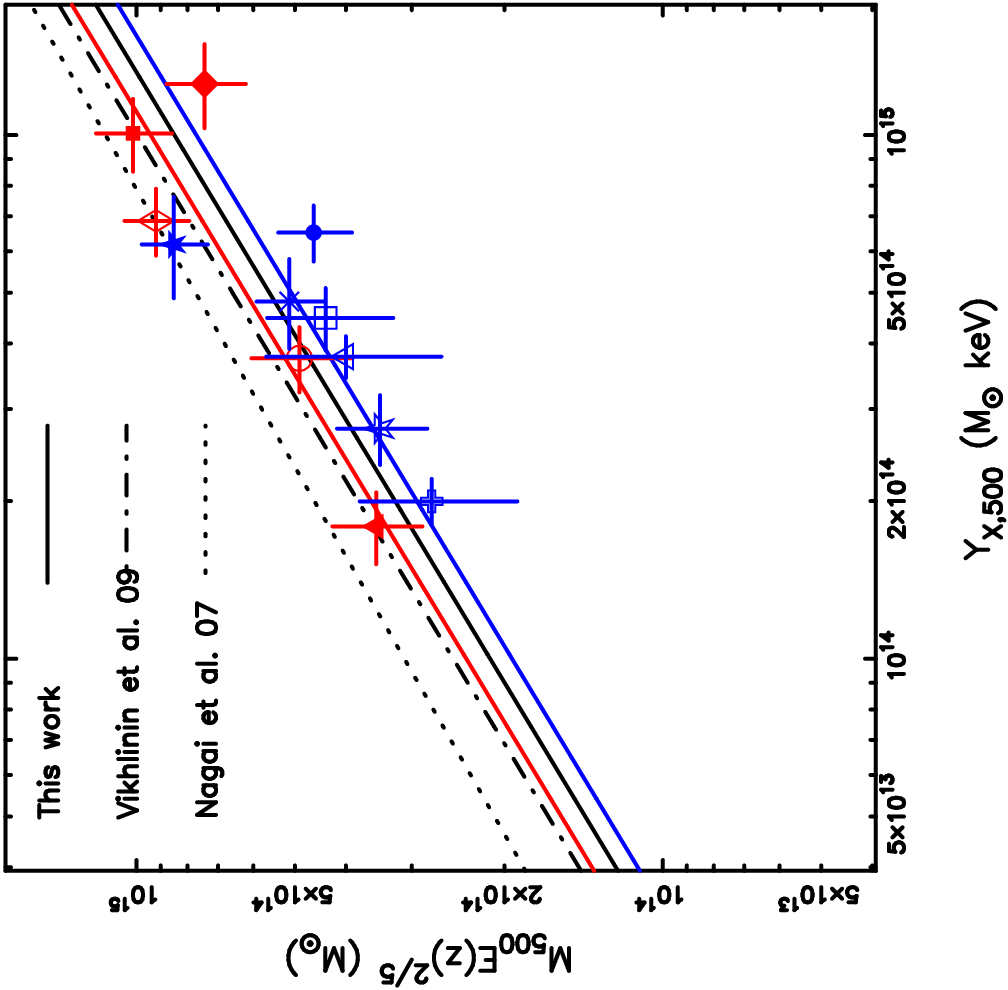}
  \caption{$M-\Yx$ relation using weak lensing masses and the
    quasi-integrated pressure at the overdensities of $\Delta=2500$
    (left panel), $\Delta=1000$ (middle panel), and $\Delta=500$
    (right panel), see also Sec. \ref{subsec:xray}.  The solid
    black, red and blue lines denote the best-fits of the
    relations using weak lensing masses for all, undisturbed,
    and disturbed clusters, respectively.}
  \label{f:my1}
\end{figure*}

\begin{figure*}
  \includegraphics[angle=270,width=5.5cm]{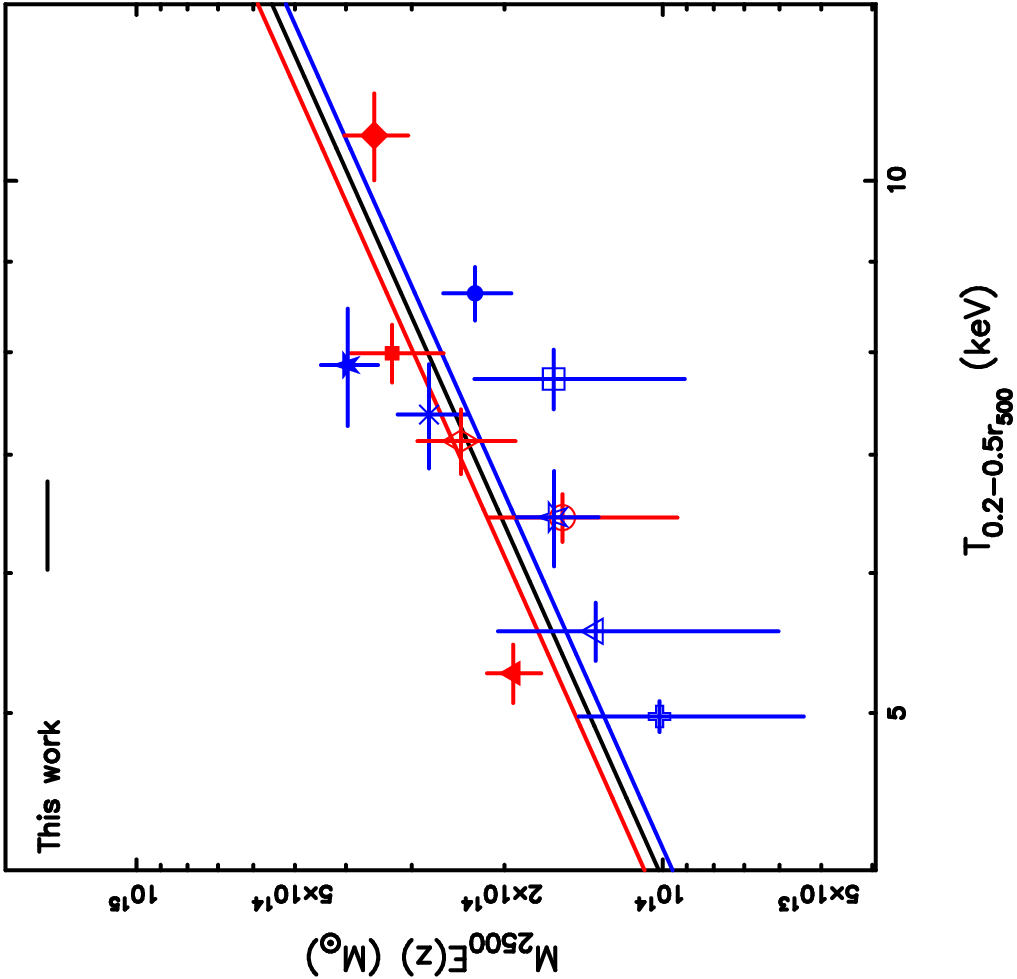}
  \hspace{-0.3cm}
  \includegraphics[angle=270,width=5.5cm]{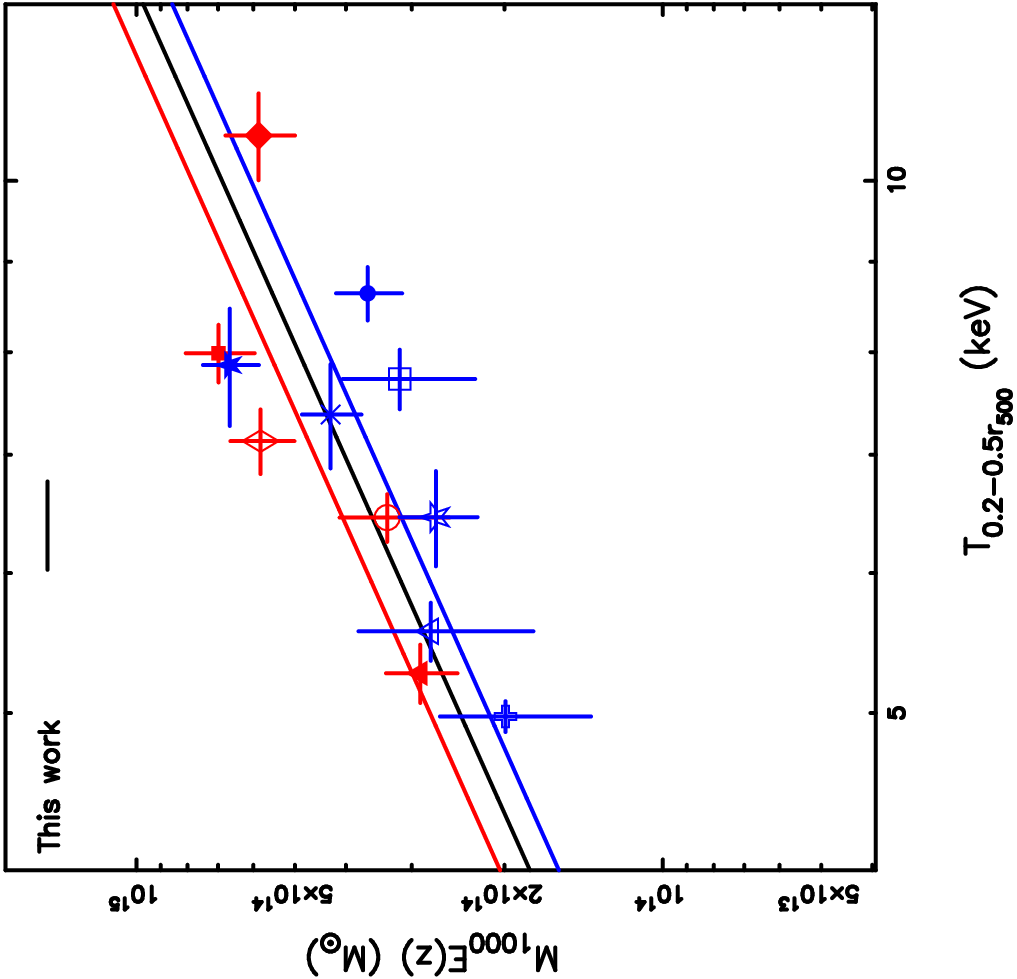}
  \hspace{-0.3cm}
  \includegraphics[angle=270,width=5.5cm]{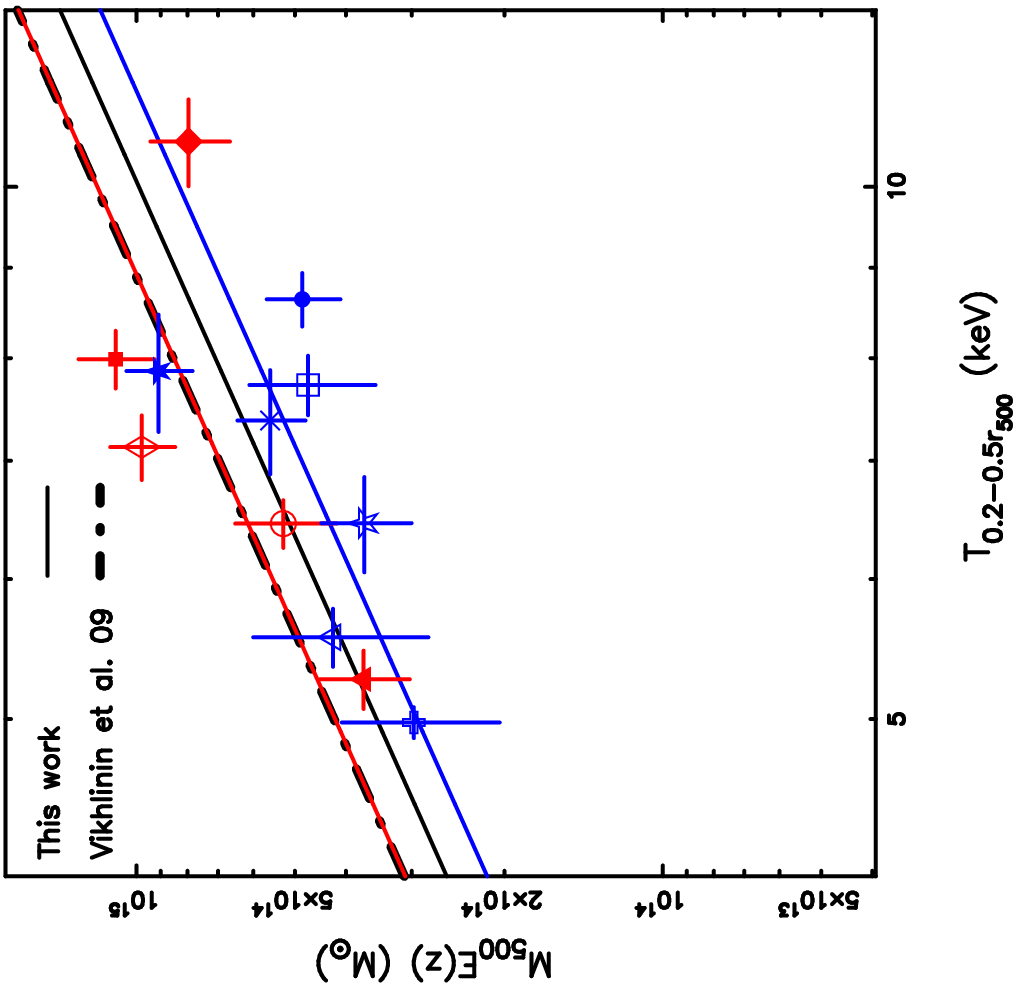}
  \caption{ $M_{\Delta}^{\rm WL}$--$T$ relation. The arrangements of
    panels with overdensity and the line coding is the same as
    Fig.~\ref{f:my1}.}
  \label{f:mt1}
\end{figure*}

\begin{figure*}
  \includegraphics[angle=270,width=5.5cm]{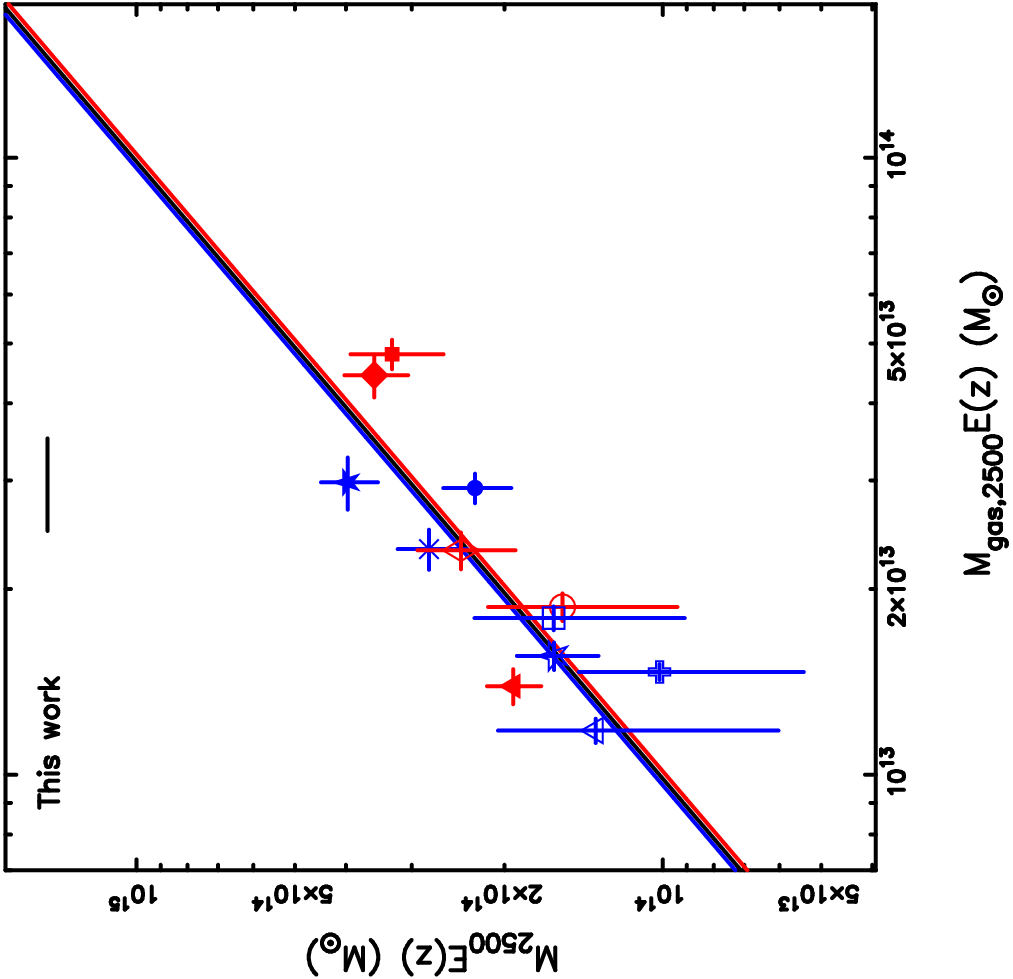}
  \hspace{-0.3cm}
  \includegraphics[angle=270,width=5.5cm]{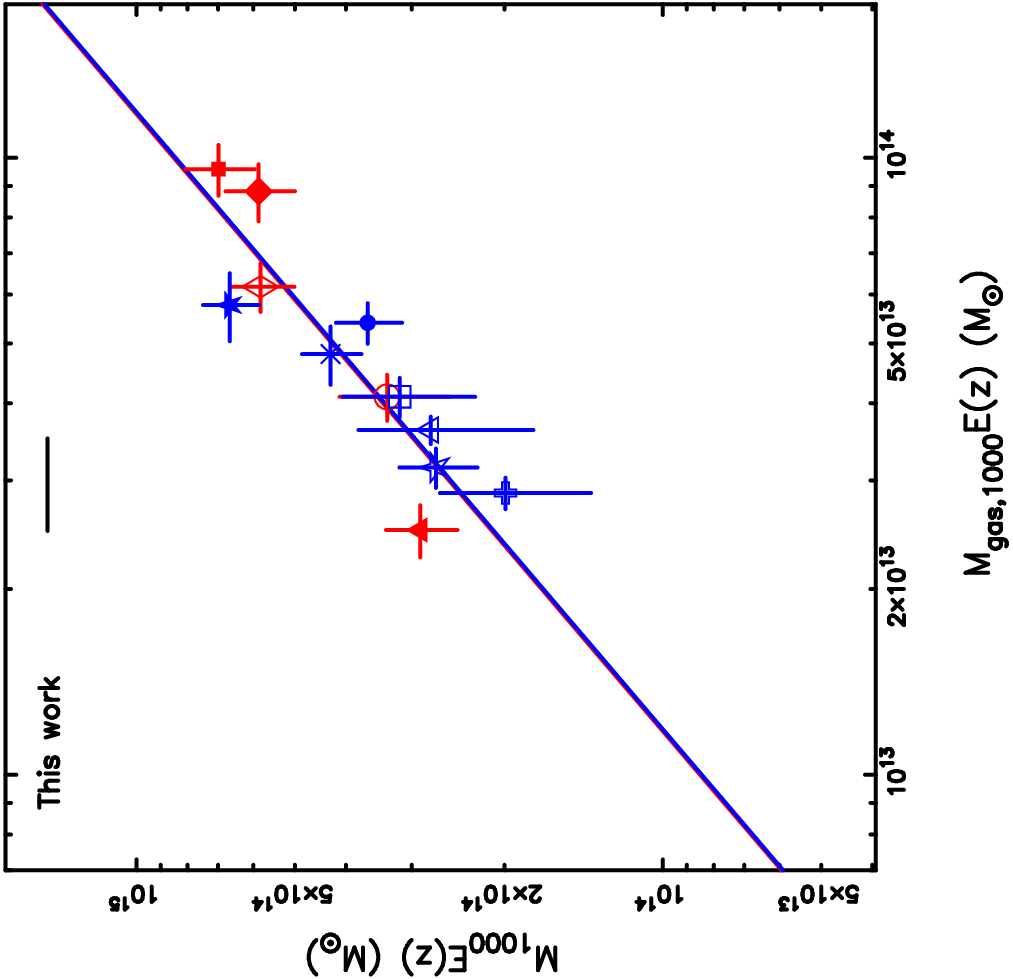}
  \hspace{-0.3cm}
  \includegraphics[angle=270,width=5.5cm]{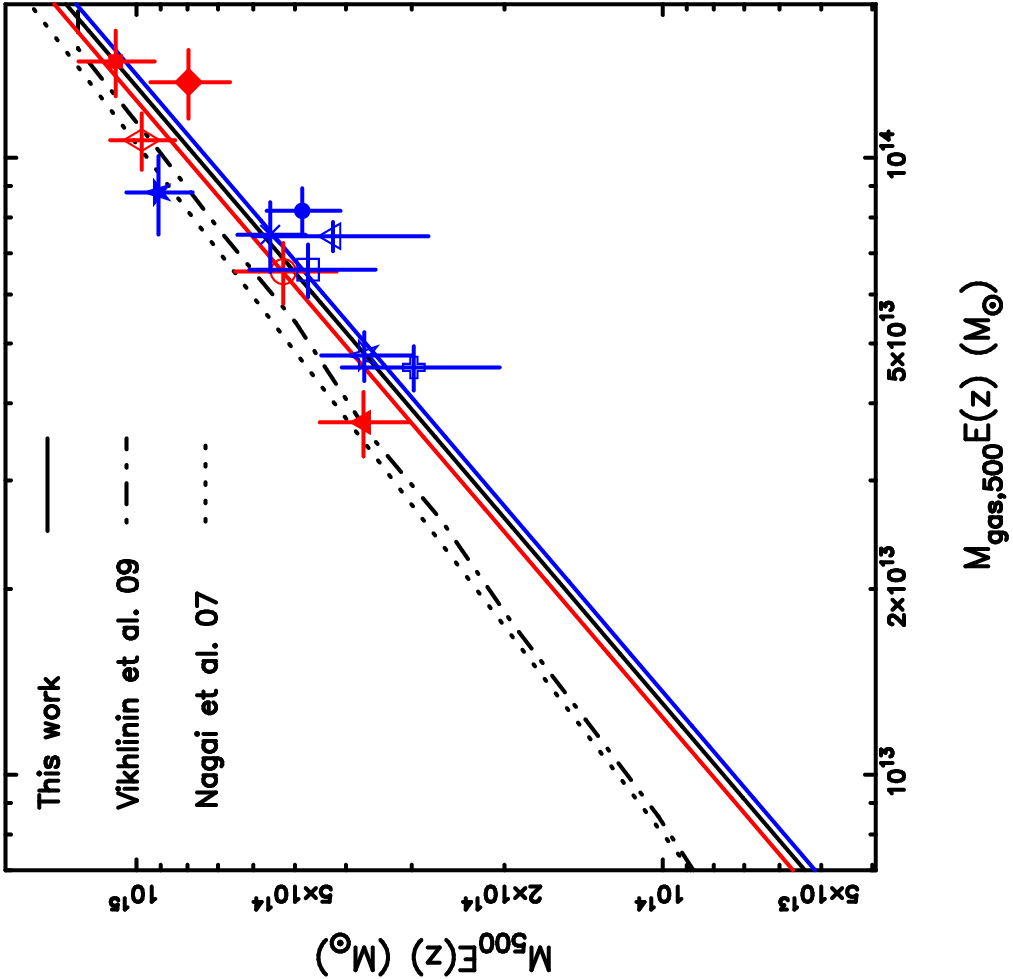}
  \caption{ $M_{\Delta}^{\rm WL}$--$\Mgas$ relation. The arrangements
    of panels with overdensity and the line coding is the same as
    Fig.~\ref{f:my1}. }
  \label{f:mmg1}
\end{figure*}

We first fit the scaling relations to the full sample of 12 clusters
with both slope $\gamma$ and normalization $M_0$ as free parameters.
At $\Delta=500$ the best-fit slopes of all three relations agree well
with the self-similar model (Table~\ref{tab:slope}).  At higher
overdensities, the agreement deteriorates for all three relations,
indeed the slopes of the $M_{2500}-M_{\rm gas}$, $M_{2500}-\Yx$
relations are discrepant from self-similar at $\sim2-3\sigma$ at
$\Delta=2500$.  This flattening in the scaling relations at higher
$\Delta$ can also be seen graphically in Figures~\ref{f:my1},
\ref{f:mt1} and \ref{f:mmg1}, in which we show the $M_\Delta$--$\Yx$,
$M_\Delta$--$T_{\rm X}$ and $M_\Delta $--$\Mgas$ relations
respectively.  


To constrain the normalization parameter $M_0$, we fix the slope
parameters to the self-similar values and repeat the fits.  The
measured normalizations are all consistent with those obtained by
Zhang et al.\ (2008) using the same \emph{XMM-Newton} data and
independent weak-lensing data.
The superior quality and uniformity of our Subaru data shows 
differences between the normalizations for disturbed and undisturbed
clusters.  
These differences are most pronounced at $\Delta=500$ --
see Table~\ref{tab:norm} -- specifically, at fixed $\Yx$ undisturbed clusters are
measured to be $\sim22\%$ more massive than disturbed clusters at
$\sim1.5\sigma$ significance.  Similarly, at fixed $T$ undisturbed
clusters are measured to be $\sim43\%$ more massive than disturbed
clusters at $\sim1.8\sigma$ significance.  We confirm that our
results are insensitive to whether or not the slopes are fixed 
to
the
self-similar value.


\subsection{Scatter}\label{subsec:scatter}

We also measured the intrinsic scatter, $\sigma_{\ln\!M}$, for the
logarithm of the $Y$-axis, $M E(z)$, for each mass-observable scaling
relation using the Bayesian method described in
the Appendix.  Here we take into account the correlations
caused by measuring X-ray observables within radii defined by the
lensing analysis -- see the Appendix.  We also confirmed
that the best-fit slopes and normalizations obtained using the ODR
methods discussed above are consistent within errors with those
obtained using the more sophisticated Bayesian method considered
here.  The intrinsic scatter in all three relations is well described
by a lognormal distribution. 

The $M-T$ relation exhibits the largest intrinsic scatter
($\sim0.23-0.33$; Table~\ref{tab:int_scat}) among the three mass-observable relations.  We also
observe an increase in the intrinsic scatter with increasing radius
(i.e.\ decreasing the interior overdensity $\Delta$ ).  The same trend
is found in undisturbed clusters, while the opposite trend is found in
disturbed clusters.  However, this trend is not a physical 
feature of the intracluster gas affected by gravitational heating,
because we used a fixed global temperature measurement in the radial
range of $0.2-0.5r_{500}$ for all the overdensities. 

The $M-\Mgas$ relation is the tightest of the three, with an intrinsic
scatter in mass of $\sigma_{\ln\!M}\sim0.12-0.16$ at
$\Delta=500$ and $1000$.  At $\Delta=2500$ the scatter is roughly
double that at lower overdensity (Table~\ref{tab:int_scat}), which may be due to different core properties of
individual clusters.  For example, cool-core clusters have denser,
cuspier cores than non-cool core clusters (e.g., Croston et al. 2008;
Sanderson et al. 2009b; McCarthy et al. 2008).  Such differences
between cluster cores have a much smaller effect on measurements at
larger radii because the core regions make a small contribution to the
total gas mass measured out to $\Delta=1000$ and $500$.  However, note
that the intrinsic scatter is not well constrained for $M-\Mgas$
because the scatter is dominated by statistical errors.  A larger
sample is clearly needed to improve the constraints on the intrinsic
scatter in $M-\Mgas$, however, it is important to note that this is the
only relation that appears to have $\sim10\%$ intrinsic scatter.

The observed intrinsic scatter in the $M-\Yx$ relation is intermediate
between that of the $M-\Mgas$ and $M-T$ relations, at
$\sigma_{\ln\!M}\sim0.20-0.25$ (Table~\ref{tab:int_scat}).  This is a factor of $\gs2$
greater than that originally predicted by Kravtsov et al.\ (2006)
based on their AMR simulations.

\subsection{The impact of an outlier} \label{subsec:outlier}

In this section we highlight the impact of one cluster, A\,1914, on
our results.  This cluster has previously been identified as a merging
cluster with a complex X-ray morphology, radio halo, and
weak-lensing-based dark matter distribution (Buote \& Tsai 1996;
Bacchi et al.\ 2003; Govoni et al.\ 2004; Okabe \& Umetsu 2008).  We
have also identified it as having the most extreme X-ray/lensing mass
discrepancy among the 12 clusters considered here (Zhang et al.\
2010).  

To assess the impact of such clusters on the measured reliability of
X-ray observables as mass proxies we repeated the calculations of
normalization and scatter discussed in Sec.~\ref{subsec:norm}~\&~\ref{subsec:scatter} excluding
A\,1914 (Table~\ref{tab:norm}).  At $\Delta=500$ the normalization of
the $M-\Yx$ relations for disturbed and undisturbed clusters are
different at just $\sim1.2\sigma$ significance when A\,1914 is
excluded from the disturbed sample, in contrast to the $\sim1.8\sigma$
difference based on the full sample of 12 clusters.  We also find that
excluding A\,1914 reduces the intrinsic scatter on all of the scaling
relations.  In particular, the intrinsic scatter on $M-\Yx$
is reduced by $\sim25\%$ from $\sigma_{\ln\!M}\sim0.20$ to $0.15$.
Jack-knife tests on samples of 11 clusters (i.e.\ removing each
cluster in turn) also confirm that A\,1914 is indeed the most
significant outlier among our sample.

These results indicate that outliers in the cluster population require
careful treatment in the construction and application of
mass-observable scaling relations.  In summary, reliable cluster
selection functions are required to gain robust constraints.  This
will be especially true for future high redshift surveys because the
fraction of merging clusters is expected to increase with look-back
time (Vikhlinin et al.\ 2009a).

\subsection{Covariance of deviations} \label{sec:scat}

We investigate the covariance of deviations from the best-fit
$M-\Mgas$ and $M-T$ relations following recent numerical simulation
studies (e.g.\ Kravtsov et al.\ 2006; Stanek et al.\ 2010).  For a
given {\em mean} scaling relation $Y=f(X)$, the deviations of each
cluster from the mean relation are quantified as
$\delta\!Y\equiv\![Y-f(X)]$ and $\delta\!X\!\equiv[X-f^{-1}(Y)]$.  We
use the mean normalizations for a full sample of 12 clusters, however
we found that the following results do not change significantly when
the best-fit normalizations of undisturbed and disturbed clusters are
used instead.

\begin{figure*}
  \includegraphics[angle=270,width=5.5cm]{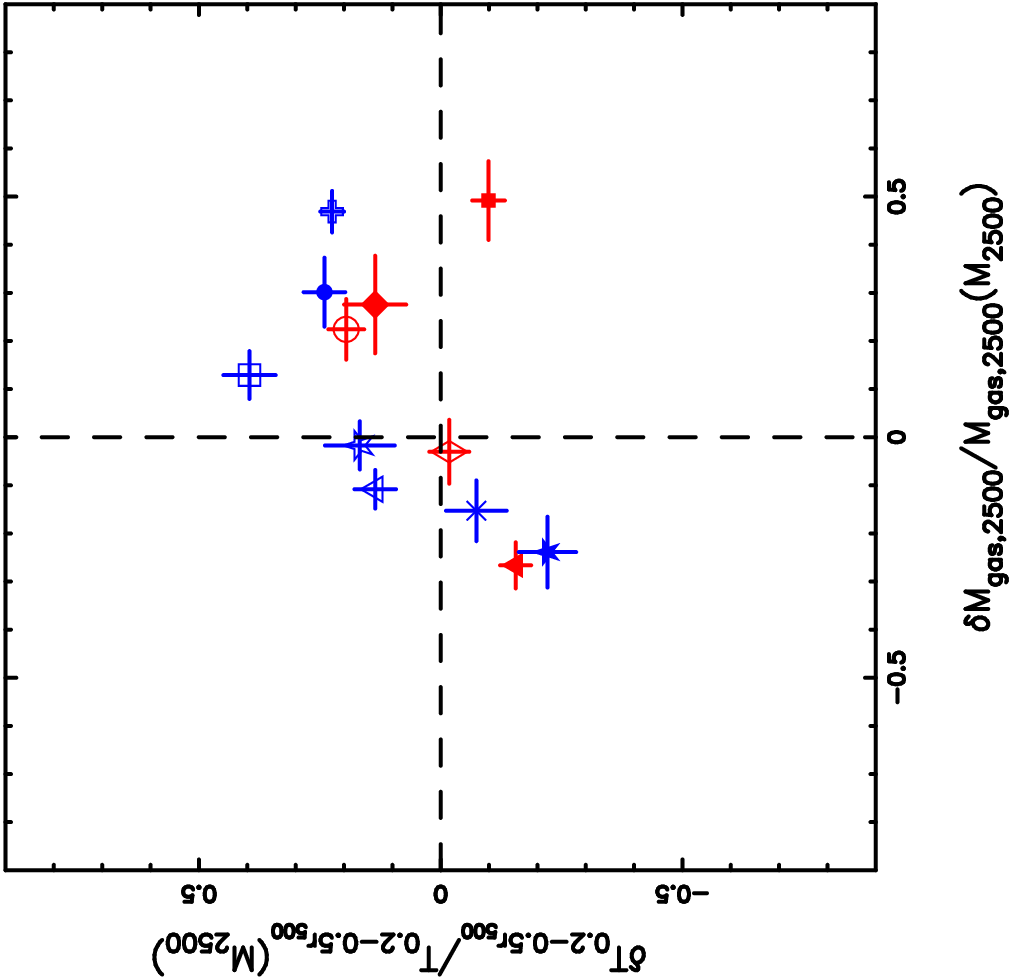}
  \hspace{-0.3cm}
  \includegraphics[angle=270,width=5.5cm]{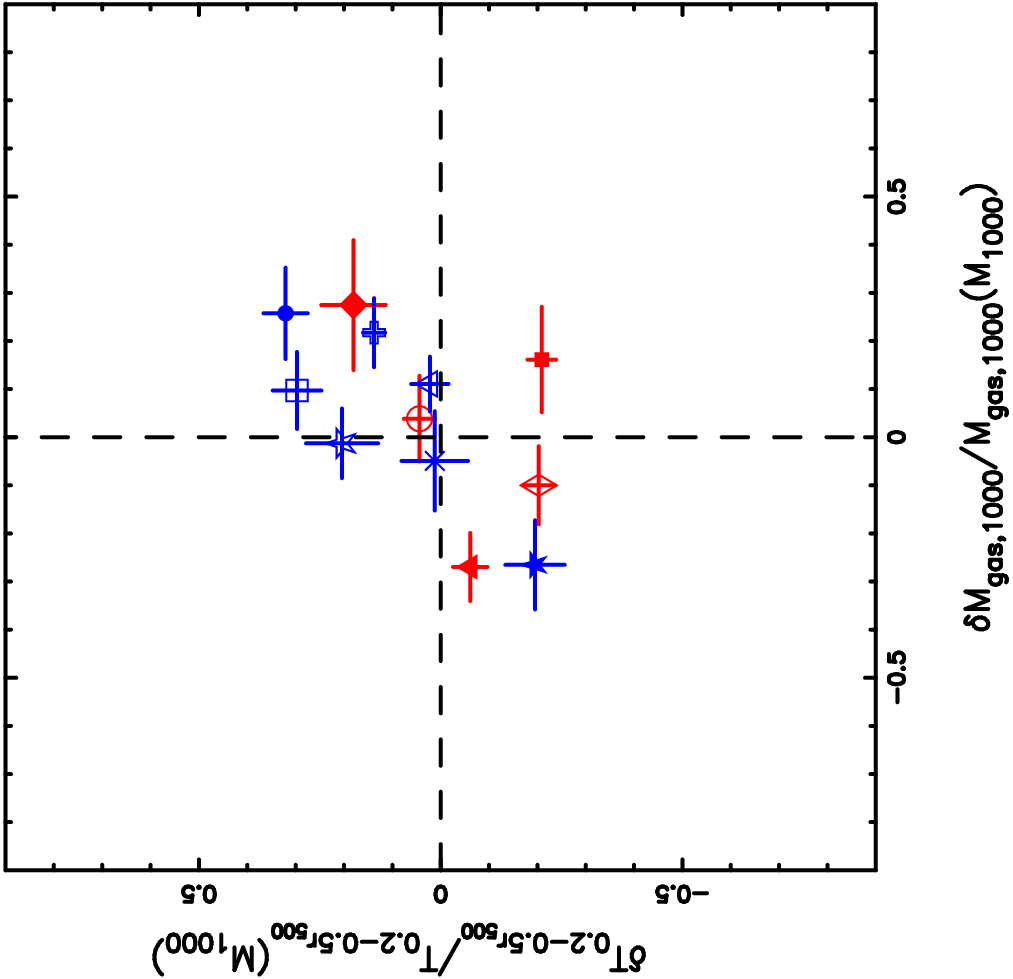}
  \hspace{-0.3cm}
  \includegraphics[angle=270,width=5.5cm]{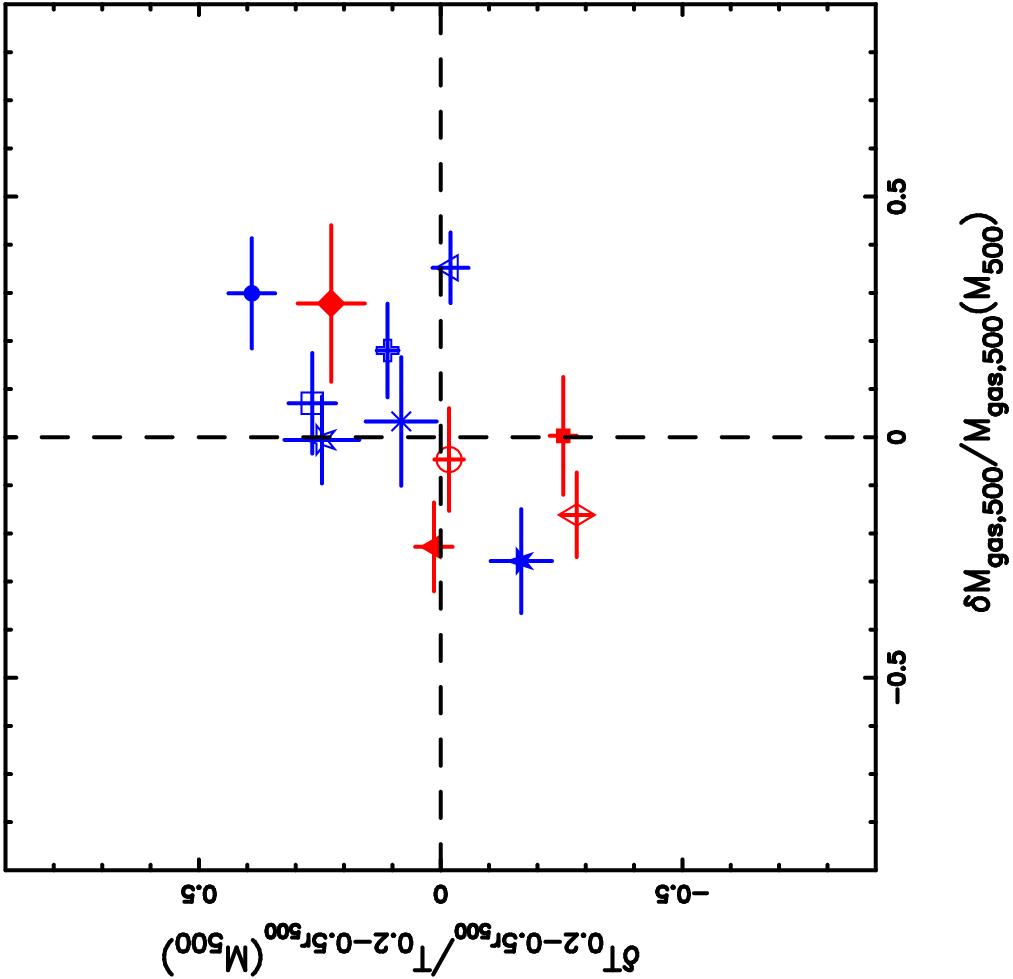}
  \caption{ Normalized temperature deviations from $M$--$T$ versus
    normalized gas mass deviations from $M$--$\Mgas$, from the
    best-fit scaling relations for the full sample at the radii with
    $\Delta=2500$ (left), $\Delta=1000$ (middle), and $\Delta=500$
    (right), respectively. The colors and symbols have the same
    meaning as in Fig.~\ref{f:my1}.}
  \label{f:dev1}
\end{figure*}

The temperature and $\Mgas$ deviations, $\delta\!T/T(M_\Delta)$ and
$\delta\!\Mgas/\Mgas(M_\Delta)$, appear to be positively correlated
(Fig.~\ref{f:dev1}).  We test this quantitatively using Spearman's
rank correlation coefficient test, obtaining $r_s=0.531\pm0.009$.  The
probability of obtaining a value of $r_s$ greater than or equal to the
measured value is low: $\mathcal{P}=0.075\pm0.006$.  This test
therefore indicates that the positive correlation is significant.
However, the apparent positive correlation between the temperature and
gas mass deviations does not show the correlation between 
intrinsic scatter, but between total scatter which is a convolution of
measurement errors and intrinsic scatter, 
because we here did not take into account for measurement uncertainties.
When dealing with observational constraints on scaling relations, it  
therefore essential to include both the  covariance of intrinsic  
scatter, and the measurement errors with which the scatter is  
convolved in robust calculations.


\subsection{Covariance of intrinsic scatter} \label{subsec:cov_int}

%

We simultaneously fit $M-T$ and $M-\Mgas$ relations and measure the covariance of the intrinsic
scatter using a multi-dimensional 
 fitting method described in the Appendix. This method considers not only the matrix of the
observational errors for individual clusters, $\mbox{\boldmath
  $\Sigma$}_{{\rm obs},i}$, but also the covariance matrix of the
intrinsic scatter, $\mbox{\boldmath $\Sigma$}_{\rm int}$.  The
covariance of the intrinsic scatter is given by
\begin{eqnarray}
\mbox{\boldmath $\Sigma$}_{\rm int} = \left(
\begin{array}{cc}
\sigma_{t}^2 & \sigma_{tg}\\
\sigma_{tg} & \sigma_{g}^2
\end{array}
\right) \label{eq:cov}
\end{eqnarray}
where $\sigma_t^2$ and $\sigma_g^2$ are variances for the logarithm of
temperature and gas mass, respectively, and
$\sigma_{tg}=r\sigma_t\sigma_g$ is a covariance with a coefficient
$r$.  Here, we do not need to take into account for intrinsic scatter on
mass, because the gas properties, under a cluster mass given by
the cosmology, only physically have intrinsic scatter due to the gas
evolution. 
When we estimate cluster masses from X-ray observables via scaling relations,
there is intrinsic scatter on mass due to the propagation from gas
intrinsic scatter.
As shown in the Appendix, the observational error
matrix for individual clusters is given by
\begin{eqnarray}
\mbox{\boldmath $\Sigma$}_{{\rm obs}} = 
 \left(
\begin{array}{cc}
   \frac{4}{9}e_{m}^2 + e_t^2  &  \frac{2}{3} (1-\frac{\partial \ln \Mgas}{\partial\ln M})e_{m}^2\\
   \frac{2}{3} (1-\frac{\partial \ln \Mgas}{\partial\ln  M})e_{m}^2~ 
   &  (1-\frac{\partial\ln \Mgas}{\partial \ln M})^2e_{m}^2+e_g^2
\end{array}
\right) \nonumber \label{eq:cov}
\end{eqnarray}
where $e_m, e_t$ and $e_g$ are observational errors for the logarithm
of the mass, temperature and gas mass, respectively.  The coefficients
of $e_m$ are the slopes of the mass observable relations $T\propto
M^{2/3}$ and $\Mgas\propto M$, and a term to account for propagation
of the error on $r_\Delta$ (derived from the lensing analysis) to the
error on the X-ray observables, $\partial \ln \Mgas/\partial\ln M$
(see the Appendix).  Since the correlation between the
observational errors of temperature and gas mass is negligible, we set
them to zero.  We measure the intrinsic scatter for the $M-\Mgas$ and
$M-T$ relations ($\sigma_g, \sigma_t$ and $r$) with their slopes set to
the self-similar prediction.  The likelihood (eq.~\ref{eq:likelihood})
is given in the Appendix.  We explore parameter space using
the Metropolis Hastings algorithm, as described in the Appendix,
restricting ranges of values explored to $0\leq\sigma_t\leq1$,
$0\leq\sigma_g\leq1$ and $|r|\leq1$.

We first impose a flat prior on $\sigma_t, \sigma_g$, and $r$ within
the limits referred to above, and obtain measurements of intrinsic
scatter that are in good agreement with those obtained by fitting the
$M-\Mgas$ and $M-T$ relations independently:
$\sigma_g=0.132_{-0.086}^{+0.066}$ and
$\sigma_t=0.213_{-0.080}^{+0.055}$ (Table~\ref{tab:Sint}).  However, the coefficient, $r$, is
not well determined due to the tail of the posterior probability
distribution extending to negative values.  Nevertheless, we derive a
$68.3\%$ confidence lower limit of $r\gs0.185$.  We also measure the
mode of the marginalized posterior probability distribution to be
$r=0.575$.

We then repeat the fit, this time using Gaussian priors centered on
the best-fit measurements of $\sigma_t$ and $\sigma_g$.  As shown in
the second column in Table \ref{tab:Sint}, all resulting parameters
are consistent with those in the flat prior.  This time the mode of
the posterior probability distribution is $r=0.570$., and we find that
the lower limit on $r$ is again positive -- i.e.\ it is positive
independent of the prior.  The positive coefficient leads to a large
intrinsic scatter of $\Yx=\Mgas T$ because the last term in
$\sigma_{\Yx}^2=\sigma_g^2+\sigma_t^2+2r\sigma_t\sigma_g$ becomes
positive.  The positive coefficient indicates that deviations in gas
mass and temperature are partially correlated.

However, we stress that the small size of our observed sample severely limits the statistical power of our results -- specifically, we
cannot rule out the possibility that the gas mass and temperature
deviations are two random variables that are correlated by accident.
To quantify this, we calculated the probability, ${\mathcal P}(|r'|\ge
|r|)$, that the correlation coefficient of the two random variables in
a sample of 12 drawings, $r'$, is higher than the observed value (Pugh
\& Winslow 1966).  Since the probability is non-negligible (${\mathcal
  P}(|r'|\ge |r|)<0.565~ and ~0.597$; Table \ref{tab:Sint}),
we cannot yet completely exclude this possibility.  Therefore, we need
to increase the sample size before we can make definitive statements
on the correlation of intrinsic scatter.

\section{Discussion}\label{sec:discuss}

\subsection{Comparison with previous observations}

First we compare our results with those of our pilot study (Zhang et
al.\ 2008).  The main difference between Zhang et al.'s analysis and
that presented here is that Zhang et al.'s weak-lensing mass
measurements were drawn from the literature, and thus suffered
heterogeneous image quality, observed depth, and systematic
uncertainties relating to background galaxy selection and faint galaxy
shape measurement.  Despite these differences, the overall
normalization of our mass-observable scaling relations agree within
the uncertainties with those of Zhang et al. (2008).  
However despite our sample being roughly a factor of two smaller than
that of Zhang et al., we detect structural segregation in the $M-\Yx$
and $M-T$ relations at $\sim2\sigma$ significance.  Our ability to
make this detection is likely due to the factor of $\gs2$ smaller
statistical errors on weak-lensing mass measurements of individual
clusters, thanks to the superb quality of our Subaru data.

This is the first time that structural segregation has been found in
the $M-\Yx$ relation, however it has been detected in the $M-T$
relation at a similar amplitude, and level of significance in previous
joint lensing/X-ray studies (Smith et al.\ 2005; Pedersen \& Dahle
2007).  Theoretical studies (e.g.\ Randall et al.\ 2002) suggest that
this segregation may be caused by cluster-cluster mergers boosting the
temperature of disturbed merging clusters.  However orientation
effects may also contribute to systematic errors in weak-lensing mass
measurements that exaggerate the segregation (Meneghetti et al.\
2010).  We therefore defer physical interpretation of the observed
segregation to a future careful investigation of the degeneracy
between cluster orientation and residuals on mass-observable scaling
relations.

We also compare the normalizations of our $M-\Yx$ and $M-T$ relations
at $\Delta=500$ with the same from Vikhlinin et al.'s (2009a)
X-ray-only study of ``relaxed'' clusters with \emph{Chandra}.  Note
that, as discussed in Sec.~\ref{sec:m-o}, our results are insensitive to whether
we use orthogonal regression or bisector fitting techniques; Vikhlinin
et al.\ used the bisector method.  The normalizations of our $M-\Yx$
and $M-T$ relations for undisturbed clusters agree within the
uncertainties with Vikhlinin et al.'s (2009a) relaxed clusters
(Figs.~\ref{f:my1}~\&~\ref{f:mt1}).  This suggests that orientation effects may not be a
major influence on the normalization of our undisturbed cluster
scaling relations.  On the other hand, it is thus clear that the
normalizations of our disturbed cluster scaling relations differ from
Vikhlinin et al.'s results at $\sim2\sigma$ significance.  It will
therefore be important in the future to study the mass-observable
scaling relations with large samples comprising clusters that span a
wide range of dynamical states.  This need will become more acute at
high redshift because the fraction of disturbed clusters likely increases
toward higher redshifts.

\subsection{Comparison with simulations}

The normalizations of our $M-\Mgas$ and $M-\Yx$ relations are lower in
mass at fixed X-ray observable (Fig.~\ref{f:my1}~\&~\ref{f:mt1}) than the predictions
from Nagai et al.'s (2007) simulations.  This result is consistent
with our pilot study (Zhang et al.\ 2008), in which we compared our
joint lensing/X-ray observational results with a wider range of
simulations including those of Borgani et al.\ (2004) 
and showed that the normalizations disagree with the simulations at
$>2\sigma$ for $M-\Yx$ and $>1\sigma$ for $M-T$ and $M-\Mgas$ relations, respectively.

The structural segregation found in our $M-T$ and $M-\Yx$ relations --
i.e.\ undisturbed clusters are $\sim40\%$ and $\sim20\%$ more massive
than disturbed clusters at fixed temperature and fixed $\Yx$
respectively at $\sim2\sigma$ significance -- disagrees with Kravtsov
et al.'s (2006) simulations, upon which they based their proposal of
$\Yx$ as a low-scatter mass proxy.  In summary, they found no
difference in the normalization of $M-\Yx$ between clusters that they
classified as ``relaxed'' and ``unrelaxed'' (roughly equivalent to our
undisturbed/disturbed classification), and that their relaxed clusters
are less massive at fixed temperature than unrelaxed clusters -- i.e.\
opposite to our observational result. 


The intrinsic scatter in the $M-T$ and $M-\Mgas$ relations at
$\Delta=500$ ($\sim20-30\%$ and $\sim10\%$ respectively) is comparable
to Kravtsov et al.\ (2006). 
However the anti-correlation between temperature and gas mass
deviations from the mean $M-T$ and $M-\Mgas$ relations predicted by
Kravtsov et al., and that has motivated much attention on $\Yx$ as a
low scatter mass proxy, is not supported by our data, although we need
to investigate this for a larger sample. 
In contrast with Kravtsov et al. (2006), 
we find that the lower limit of the positive coefficient for intrinsic
scatter is in better agreement with Stanek et al.'s SPH Millennium Gas
Simulations showing that the coefficients between the
spectroscopic-like temperature and gas fraction for $z=0$ and
$\Delta=200$ are positive irrespective of the process of preheating and
cooling. 

Correlated gas mass and temperature deviations 
imply a possibility
that adiabatic compression/expansion of the intracluster gas is
important in cluster evolution.  These adiabatic fluctuations
propagate much faster than cooling losses, and thus help to explain
why Stanek et al.'s simple hydrodynamical simulation matches the
observations well. 
Umetsu et al.\ (2010) have also suggested that, in the adiabatic expansion phase of a post-merger,
both temperature and (encompassed) gas mass decrease compared to those
before the merger. 
Another possibility can be due to the departure from the spherical 
symmetry. As long as the intracluster gas is in H.E.,
even in the elongated gravitational potential well, the gas mass distribution 
is supposed to be more round than the distribution of dark matter. 
In our modeling, we assume spherical symmetry of the gas and mass 
profiles, in which deviations from spherical geometry are transferred 
into systematic measurement and may in part lead to a positive correlation 
between the gas mass and mass.
Furthermore, many possible deviations in
mass-observable relations could be relevant to the interpretation of
these results.  For example, star-formation efficiency affects the
total gas mass and the epoch for cluster formation affects the cluster
temperature.  Some of these effects imply other correlations that can
be explored in the future with a larger sample, e.g.\ correlations
between temperature deviations and dark matter profile shapes, or
between deviations in the gas mass and total stellar fraction.


\subsection{Principal component analysis of observational data}

Finally, we propose a method, based on principal component analysis,
for constructing and calibrate a low-scatter mass proxy using solely
observational data.  Thanks to multi-dimensional fitting, since we obtained
both the intrinsic covariance and the normalization, we here do not need
to take into account for measurement errors. 
By solving the eigenvalue equation
$(\mbox{\boldmath $\Sigma$} -\sigma^2 \mbox{\boldmath $I$})
\mbox{\boldmath $y$}=0$, where $\mbox{\boldmath $I$}$ is the identity
matrix, we can obtain its eigenvalues $\sigma_\pm^2$ and eigenvectors
$y_\pm=\ln Y_\pm$, as follows:
\begin{eqnarray}
  \sigma_{\pm}^2&=&\frac{1}{2}\left[\sigma_t^2+\sigma_g^2\pm
			     \left((\sigma_t^2-\sigma_g^2)^2 + 4 r^2\sigma_t^2\sigma_g^2
			     \right)^{1/2}\right], \\
  Y_{\pm} &=& \Mgas T^{p_\pm}, \\
  p_\pm   &=& \frac{\sigma_t^2-\sigma_g^2\pm\left((\sigma_t^2-\sigma_g^2)^2 + 4 r^2\sigma_t^2\sigma_g^2
			     \right)^{1/2}}{2 r \sigma_t\sigma_g}
\end{eqnarray}
where $Y_-$ is the mass proxy with the smaller scatter $\sigma_-$.
Note that if the coefficient $r$ is negative, as
predicted by Kravtsov et al.\ (2006), then the temperature exponent,
$p_-$, is always positive.  In this framework, the scaling relation
can be written as:
\begin{eqnarray}
 M E(z) \propto \left(Y_{\pm} E(z)\right)^{3/(3+2p_\pm)}h^{(9-4p_\pm)/(3+2p_\pm)/2},
\end{eqnarray}
and the intrinsic scatter on mass when one estimate from X-ray
observables via new scaling relation is given by:
\begin{eqnarray}
\sigma_{\ln M}=\frac{3}{|3+2p_\pm|}\sigma_\pm.
\end{eqnarray}
Basically, a combination of highly correlated/anti-correlated
observables gives a new mass proxy with smaller scatter.
Unfortunately, our sample is too small to constrain the
covariance of the intrinsic scatter well, and to perform this principal
component analysis.  This exercise awaits enlargement of our sample.
In principle, a combination of the principal component analysis and
the method for measuring the covariance of intrinsic scatter (see the Appendix) could be applied to multi-dimensional (i.e.\ weak
lensing, X-ray, SZ and optical observables) data sets both within
LoCuSS and in large forthcoming surveys.

\section{Summary} \label{sec:sum}

We have presented a joint weak-lensing and X-ray analysis of 12
clusters, based on Subaru and \emph{XMM-Newton} observations as part
of the LoCuSS Survery.  The main goal is
to calibrate the scaling relations between cluster mass obtained from
weak lensing observations ($M$), and X-ray observables, specifically
the gas temperature, gas mass, and quasi-integrated pressure ($T$,
$\Mgas$, and $Y_{\rm X}=\Mgas\times T$).  An accurate understanding of
these relations will be essential to the success of future attempts to
constrain dark energy with clusters via growth of structure
experiments.  Our main results are summarized below.

\begin{itemize}

\item The dynamical state of clusters can be diagnosed empirically via
  morphological classification of clusters based on asymmetry ($A$)
  and fluctuation ($F$) parameters derived from imaging data from
  \emph{XMM-Newton}.  Undisturbed clusters are identified as those
  with relatively symmetric and smooth X-ray morphology -- $A<1.1$ and
  $F<0.05$; disturbed clusters satisfy either or both of these
  criteria.  Five clusters are classified as undisturbed and seven as
  disturbed.  This classification matches those based on
  alternative measures such as the presence/absence of cool cores,
  cold fronts, and substructures in lensing mass maps.

\item We detected structural segregation in the $M-T$ and $M-\Yx$
  relations at $\Delta=500$, in the sense that undisturbed clusters
  are $\sim40\%$ and $\sim20\%$ more massive than disturbed clusters
  at fixed $T$ and $\Yx$ respectively, at $\sim2\sigma$ significance.
  Segregation in the $M-T$ plane is qualitatively in agreement with
  some of the previous observational results (Smith et al.\ 2005;
  Pedersen \& Dahle 2007); as far as we know these are the first joint
  lensing/X-ray results on the $M-\Yx$ relation.  These results both
  contradict Kravtsov et al.'s predictions upon which they based their
  proposal that $\Yx$ may be a useful low-scatter mass proxy.

\item The intrinsic scatter in the observed $M-T$, $M-\Yx$ and
  $M-\Mgas$ relations is measured to be $\sigma_{\ln\!M}\sim0.3$,
  $\sim0.2$, and $\sim0.1$ respectively at $\Delta=500$.
  $\Mgas$ therefore appears to be the most promising mass proxy of
  these three observables, especially because the scatter in $M-\Mgas$
  appears to be independent of the cluster dynamical state.

\item The best-fit mass-observable scaling relations are sensitive to
  the inclusion/exclusion of one cluster in our sample, namely
  A\,1914, a well known merging cluster.  If this cluster is excluded
  from our analysis then the scatter is greatly reduced -- most
  notably, the intrinsic scatter in the $M-\Yx$ relation is reduced by
  $\sim25\%$.  We conclude that a larger, more complete sample of
  clusters is required to reliably calibrate the scaling relations and
  to robustly measure how the most extreme merging clusters influence
  the relations.  This will be particularly valuable as scaling
  relation studies proceed to higher redshifts at which merging
  clusters are expected to become more prevalent.

\item Temperature deviations from the best-fit $M-T$ relation and gas
  mass deviations from the $M-\Mgas$ relation are positively
correlated.  The coefficient between the gas mass and temperature
deviations is positive, independent of our analysis methods, and is
found to be $r\ge0.185$.  This result, in particular the lower limit
on $r$ agrees well with predictions based on Millennium Gas Simulations
(Stanek et al.\ 2010), and disagrees with predictions based on
Kravtsov et al.'s (2006) simulations.  However, we caution that the
chance probability, ${\mathcal P}(|r'|\ge |r|)$, that the correlation
coefficient of the two random variables in a sample of 12 drawings,
$r'$, is higher than the observed one, is not small.  We therefore
cannot exclude the possibility that the measured correlation between
gas mass and temperature residuals is an accident.  A larger sample of
clusters is needed to achieve definitive results.

\item Finally, we outlined a new method for constructing a robust
  low-scatter mass proxy, $Y_-=\Mgas T^{p_-}$, calibrated solely by
  observational data, based on a principal component analysis.  This
  is a generalization of the quasi-integrated pressure $\Yx=\Mgas T$,
  proposed by Kravtsov et al.\ (2006).
   In principle, a combination of the principal component analysis
  and the method for measuring the covariance of intrinsic scatter 
 could be applied to multi-dimensional data sets in order to construct
 robust new mass proxy.

\end{itemize}

Our future program will concentrate on expanding the sample of
clusters for which high quality Subaru and \emph{XMM-Newton} data are
available, in order to achieve definitive results on the issues raised
in this article.  Key issues will include improvement of the
statistical uncertainties on the scaling relation fits, characterizing
more fully the influence of extreme merging clusters on the scaling
relations, and exploring the balance between physical and
orientation effects in causing the observed structural segregation in
the $M-T$ and $M-\Yx$ relations.

\section*{Acknowledgments}

We are very grateful to the members of LoCuSS collaboration, in
particular, Arif Babul, for invaluable discussions and comments.  
NO acknowledges Yuji Chinone for helpful comments on MCMC.  NO, MT and TF
are supported in part by a Grant-in-Aid from the Ministry of
Education, Culture, Sports, Science, and Technology of Japan (NO:
20740099; MT: 20740119; TF: 20540245).  YYZ and NO acknowledge support
by the DFG through Emmy Noether Research Grant RE~1462/2, through
Schwerpunkt Program 1177, and through project B6 ``Gravitational
Lensing and X-ray Emission by Non-Linear Structures'' of Transregional
Collaborative Research Centre TRR 33 ``The Dark Universe'', and
support by the German BMBF through the Verbundforschung under grant
No.\,50\,OR\,0601 and No.\,50\,OR\,1005.  AF acknowledges support from
BMBF/DLR under grant No.\,50\,OR\,0207 and MPG.  AF was partially
supported by a NASA grant NNX08AX46G to UMBC.  KU is partially
supported by the National Science Council of Taiwan under the grant
NSC95-2112-M-001-074-MY2.  GPS acknowledges support from the Royal
Society.  This work is supported by a Grant-in-Aid for the COE Program
``Exploring New Science by Bridging Particle-Matter Hierarchy'' and
G-COE Program ``Weaving Science Web beyond Particle-Matter Hierarchy''
in Tohoku University, funded by the Ministry of Education, Science,
Sports and Culture of Japan.  This work is in part supported by a
Grant-in-Aid for Science Research in a Priority Area "Probing the Dark
Energy through an Extremely Wide and Deep Survey with Subaru
Telescope" (18072001) from the Ministry of Education, Culture, Sports,
Science, and Technology of Japan. YYZ and AF acknowledge the
hospitality of the Tohoku University during their frequent visits.
This work is supported in part by World Premier
International Research Center Initiative (WPI Initiative), MEXT, Japan.

\appendix

\section{Multi-dimensional Fitting of Data with covariance of intrinsic
  scatter} \label{sec:app} 

The measurement of a covariance of intrinsic scatter beyond
observational errors is of prime importance in order to understand their
intrinsic characteristics.  We derive the multi-dimensional fitting
with the covariance of intrinsic scatter, in the context of the Markov
Chain Monte Carlo (MCMC) method with standard Metropolis-Hastings
sampling, taking into account observational errors.

Here, we suppose a data set ($\mbox{\boldmath $x$}=\{ x_i, y_{1i},
y_{2i},\ldots,y_{pi}\}_{i=0}^{n}$) of $n$ sampling numbers and $p+1$
variables.  A linear regression equation defined by $a_p + b_p x_i$
and an intrinsic covariance matrix $\mbox{\boldmath $C$}_{\rm int}$
($p\times p$) in the $y$-coordinates are applied to modeling the
relationship between $x$ and $y_p$ in the data set.  The diagonal
elements in the covariance of intrinsic scatter, $\mbox{\boldmath
  $C$}_{\rm int}$ are intrinsic scatter, $\sigma_p^2$, for the $y_p$
observables.  The off-diagonal elements describe the intrinsic
covariance between $y_p$ variables.

A fitting parameter $\mbox{\boldmath $\theta$}$ is composed of the
linear regression parameters, $\mbox{\boldmath $a$},\mbox{\boldmath
  $b$}$, and the elements of the intrinsic covariance matrix.  From
Bayesian statistics, the posterior probability of the parameter vector
$\mbox{\boldmath $\theta$}$ is proportional to the conditional
probability of $\mbox{\boldmath $x$}$, $p(\mbox{\boldmath
  $x$}|\mbox{\boldmath $\theta$})$ and a prior probability function,
$p_{\rm prior}(\mbox{\boldmath $\theta$})$,
\begin{eqnarray}
p(\mbox{\boldmath $\theta$} | \mbox{\boldmath $x$})\propto p(\mbox{\boldmath $x$}|\mbox{\boldmath $\theta$})p_{\rm prior}(\mbox{\boldmath $\theta$}).
\end{eqnarray}
The conditional probability of $\mbox{\boldmath $x$}$, given
parameters $\mbox{\boldmath $\theta$}$, is the likelihood described by
\begin{eqnarray}
p(\mbox{\boldmath $x$}|\mbox{\boldmath $\theta$}) = \prod_{i=0}^n
 \frac{1}{ (2\pi)^{p/2} |{\rm det(}\mbox{\boldmath$C_i$})|^{1/2}}\exp\left[-(\mbox{\boldmath$y_i$}-(\mbox{\boldmath$a$}+\mbox{\boldmath$b$}x))^{T}
     \mbox{\boldmath$C_i$}^{-1}(\mbox{\boldmath$y_i$}-(\mbox{\boldmath$a$}+\mbox{\boldmath$b$}x))/2\right], \label{eq:likelihood}
\end{eqnarray}
where the covariance matrix $\mbox{\boldmath $C_i$} = \mbox{\boldmath
  $C_i$}_{\rm, obs} + \mbox{\boldmath $C$}_{\rm int}$ with the
observational error covariance matrix, $\mbox{\boldmath $C$}_{\rm
  obs}$ and intrinsic covariance.  The loglikelihood of
eq. (\ref{eq:likelihood}) is given by
\begin{eqnarray}
- 2 L = \sum_i\log ({\rm det (\mbox{\boldmath $C_i$})})
 + \sum_i (\mbox{\boldmath$y_i$}-(\mbox{\boldmath$a$}+\mbox{\boldmath$b$}x))^{T}
     \mbox{\boldmath$C_i$}^{-1}(\mbox{\boldmath$y_i$}-(\mbox{\boldmath$a$}+\mbox{\boldmath$b$}x)). \label{eq:L}
\end{eqnarray}
The second term on the right-hand side of Equation (\ref{eq:L}) is referred to as the
$\chi^2$ in some papers (e.g. Akritas \& Bershady 1996; Tremaine et
al. 2002; Pizagno et al. 2005; Novak et al. 2006; Weiner et al. 2006),
and the intrinsic scatter is estimated by requiring the reduced
$\chi^2$ to be unity, in the framework of the $\chi^2$
minimization fitting.  It is, however, an inadequate fitting procedure
because the first term, depending on the parameters, cannot be ignored
(see also D'Agostini 2005).  We therefore employ the likelihood
function for a calculation of the covariance of intrinsic scatter.
The diagonal elements in the observed error matrix are given by
$(b_p-\partial y_{p,i} / \partial x_i)^2 e_{x,i}^2+e_{y_p,i}^2$ with the
observed variances $e_x^2$ and $e_{y_p}^2$ in $x$ and $y_p(x)$
variables. Here, $\partial y_i / \partial x_i$ represents the error
propagation between two variables.  The off-diagonal elements,
$C_{pp',i}$, for $i$ sample are  $(b_p-\partial y_{p,i} / \partial
x_i)(b_{p'}-\partial y_{p',i} / \partial
x_i)e_{x,i}^2+\langle e_{y_p,i} e_{y_{p'},i} \rangle$. 
It describes the correlation between observational errors : the first
term is an error correlation via the same $x$ values, and the
second one is the correlation between observational errors of $y_p$ and $y_{p'}$.
In this study, we measure the covariance of intrinsic scatter in the
parameter plane of $x=\ln(M)$, $\mbox{\boldmath $y$}=(\ln(T),
\ln(\Mgas))$.

\begin{figure*}
  \includegraphics[angle=270,width=5.5cm]{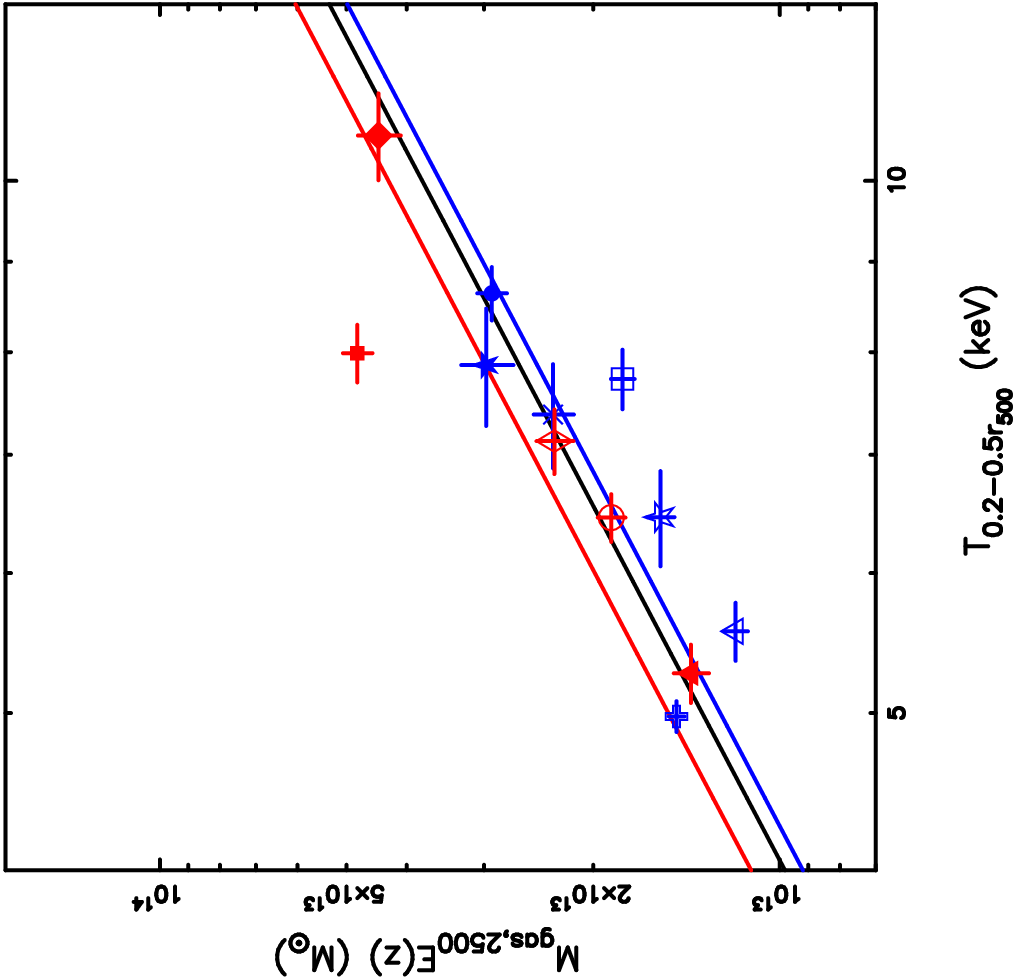}
  \hspace{-0.3cm}
  \includegraphics[angle=270,width=5.5cm]{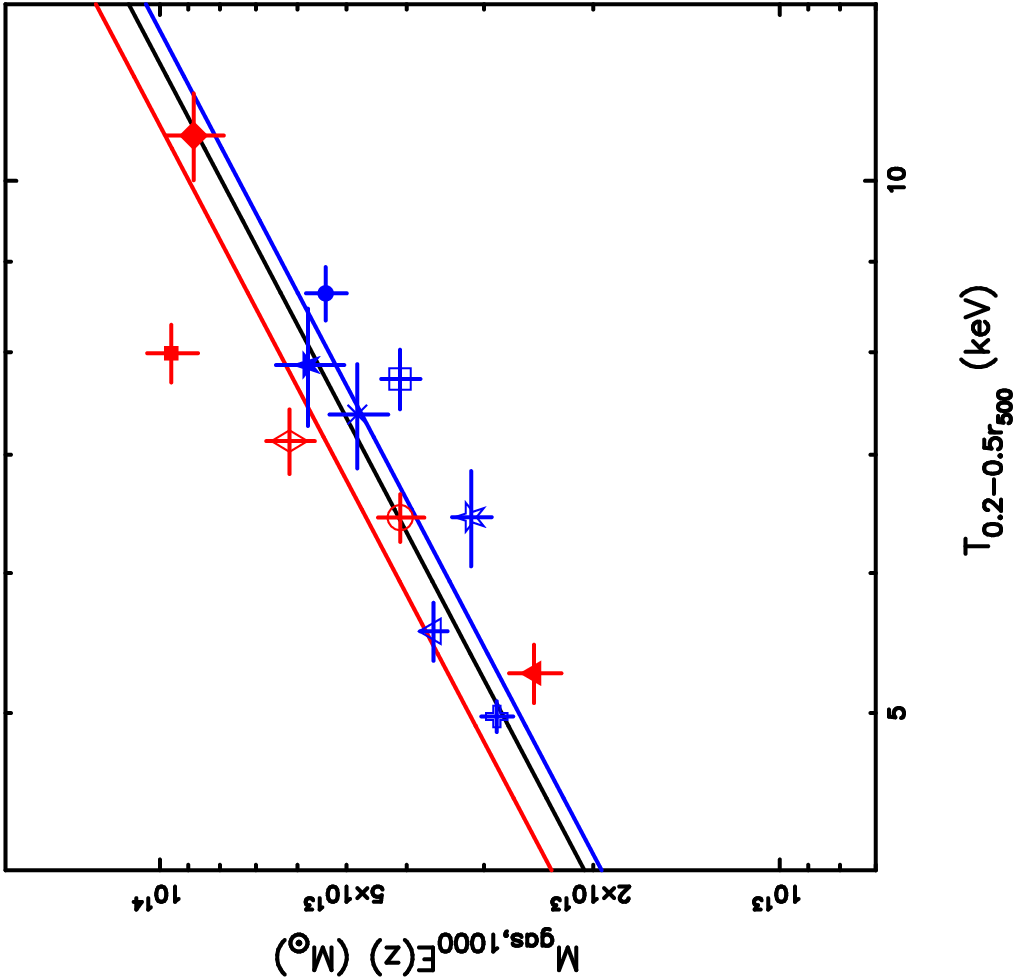}
  \hspace{-0.3cm}
  \includegraphics[angle=270,width=5.5cm]{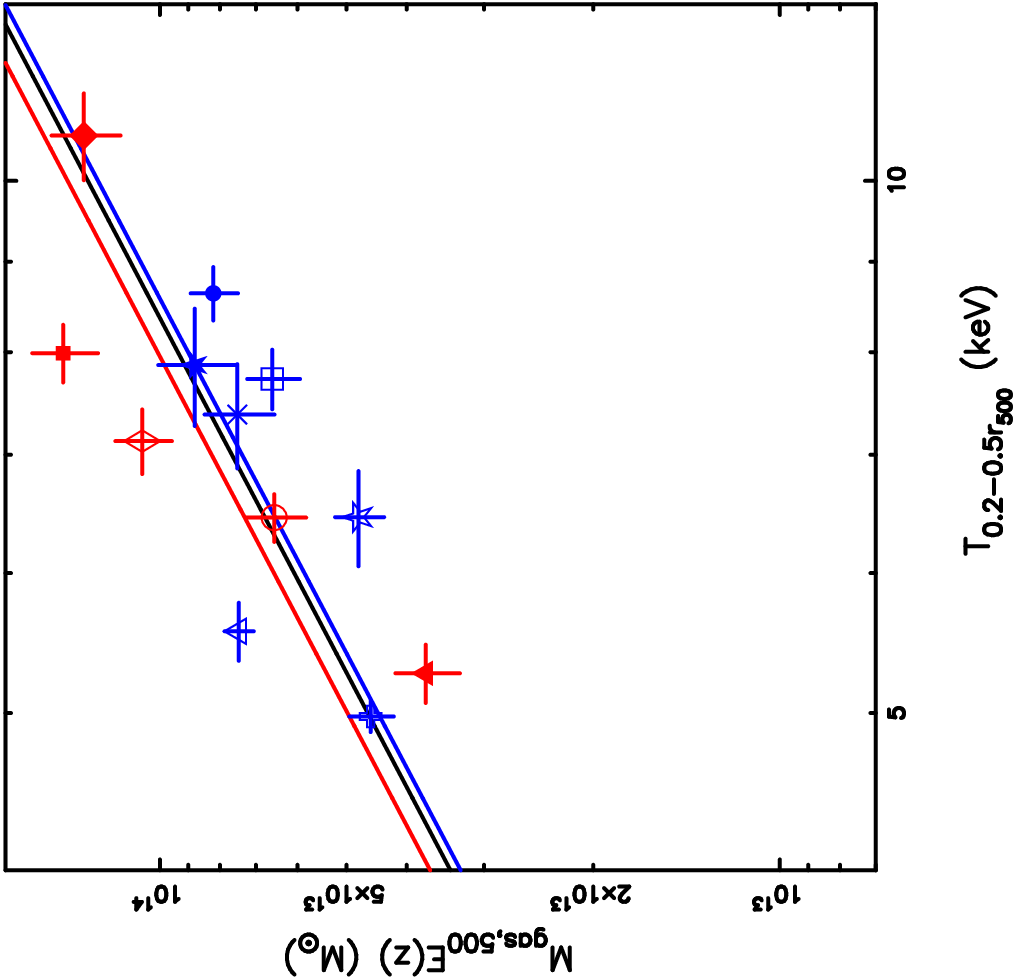}
  \caption{ $M_{\rm gas,\Delta}$--$T$ relation. The arrangements of
    panels with overdensity and the line coding is the same as
    Fig.~\ref{f:my1}. }
  \label{f:mgt1}
\end{figure*}

\begin{table*}
  \caption{Slopes of the mass-observable relations for the full
    sample.} \label{tab:slope} 
  \begin{center}
    \begin{tabular}{l||rrr}
      \hline
      \hline
      Relation  &   \multicolumn{3}{c}{$\gamma$ using $M_{\rm WL}$}\\      
      &   $\Delta$ : $500$
      &   $1000$
      &   $2500$ \\
      (1)       &   (2)
      &   (3)
      &   (4) \\
      
      \hline
      $M_{\Delta}$--$\Yx^\gamma$
      & $0.67\pm0.14$
      & $0.59\pm0.11$
      & $0.46\pm0.11$ \\
      \hline
      $M_{\Delta}$--$T_{\rm 0.2-0.5r_{500}}^\gamma$
      & $1.49\pm0.58$
      & $1.49\pm0.46$
      & $1.26\pm0.35$ \\
      \hline
      $M_{\Delta}$--$\Mgas^\gamma$
      &  $0.98\pm0.15$
      &  $0.86\pm0.14$
      & $0.68\pm0.15$ \\
      \hline
    \end{tabular}
  \end{center}
  \tablecomments{A single power law form of slope $\gamma$ is considered. 
    The X-ray temperature is derived by the volume average of the deprojected
    radial temperature profile and using the radii within
    $0.2-0.5r_{\rm 500}$ ($T=T_{\rm 0.2-0.5r_{500}}$; see Zhang et
    al. 2008). 
    Column (1): Scaling relations.
    Columns (2-4): Slopes of the mass-observable
    relations using weak lensing masses at the overdensities of
    $\Delta=500,~1000~\&~2500$, respectively. }
\end{table*}

\begin{table*}
  \caption{Normalization ($M_0$) and morphological dependence of the
    mass-observable relations. } \label{tab:norm} 
  \begin{center}
    \begin{tabular}{l||rrr|rr}
      \hline
      \hline
      Relation  &  All 12 clusters
      &
      &
      &  w/o A1914
      & \\
      &   all
      &   undisturbed
      &   disturbed
      &   all
      &   disturbed \\
      (1)       &   (2)
      &   (3)
      &   (4)
      &   (5)
      &   (6) \\
      \hline
      $M_{500}$--$\Yx$ 
      & $4.92^{+0.35}_{-0.33}$
      & $5.46^{+0.57}_{-0.52}$
      & $4.47^{+0.41}_{-0.37}$
      & $5.16^{+0.34}_{-0.32}$
      & $4.83^{+0.41}_{-0.38}$ \\
      $M_{1000}$--$\Yx$ 
      & $4.51^{+0.29}_{-0.27}$
      & $4.77^{+0.43}_{-0.40}$
      & $4.28^{+0.39}_{-0.36}$
      & $4.70^{+0.28}_{-0.26}$
      & $4.62^{+0.41}_{-0.37}$ \\
      $M_{2500}$--$\Yx$   
      & $4.02^{+0.29}_{-0.27}$
      & $4.06^{+0.49}_{-0.43}$
      & $3.99^{+0.41}_{-0.37}$
      & $4.20^{+0.30}_{-0.28}$
      & $4.38^{+0.43}_{-0.39}$ \\
      \hline
      $M_{500}$--$T_{\rm 0.2-0.5r_{500}}$
      & $2.45^{+0.27}_{-0.24}$
      & $2.94^{+0.50}_{-0.42}$
      & $2.05^{+0.24}_{-0.22}$
      & $2.34^{+0.13}_{-0.12}$
      & $2.20^{+0.10}_{-0.10}$  \\
      $M_{1000}$--$T_{\rm 0.2-0.5r_{500}}$
      & $1.70^{+0.15}_{-0.14}$
      & $1.94^{+0.24}_{-0.21}$
      & $1.50^{+0.18}_{-0.16}$
      & $1.80^{+0.15}_{-0.14}$
      & $1.64^{+0.19}_{-0.17}$ \\
      $M_{2500}$--$T_{\rm 0.2-0.5r_{500}}$
      & $0.97^{+0.08}_{-0.07}$
      & $1.03^{+0.10}_{-0.10}$
      & $0.91^{+0.12}_{-0.10}$
      & $1.02^{+0.08}_{-0.07}$
      & $1.00^{+0.14}_{-0.12}$ \\
      \hline
      $M_{500}$--$\Mgas$
      & $13.10^{+0.77}_{-0.73}$
      & $13.79^{+1.23}_{-1.13}$
      & $12.53^{+1.03}_{-0.96}$
      & $13.59^{+0.77}_{-0.73}$
      & $13.36^{+1.09}_{-1.01}$\\
      $M_{1000}$--$\Mgas$
      & $14.46^{+0.86}_{-0.81}$
      & $14.52^{+1.52}_{-1.38}$
      & $14.41^{+1.13}_{-1.05}$
      & $14.91^{+0.89}_{-0.84}$
      & $15.36^{+1.14}_{-1.06}$\\
      $M_{2500}$--$\Mgas$
      & $17.33^{+1.36}_{-1.26}$
      & $16.85^{+2.54}_{-2.21}$
      & $17.76^{+1.65}_{-1.51}$
      & $18.01^{+1.46}_{-1.35}$
      & $19.45^{+1.51}_{-1.40}$ \\
      \hline
    \end{tabular}
  \end{center}
  \tablecomments{The forms of
    $M$--$\Yx$, $M$--$T_{\rm X}$, and $M$--$\Mgas$ relations are
    given by $M_\Delta E(z)^{2/5}= M_0(Y_{\rm x}/3\times10^{14}M_\odot {\rm
      keV})^{5/3}\times10^{14}h^{1/2}M_\odot$, $M_\Delta E(z)=M_0(k_B T/5~{\rm
      keV})^{3/2}\times10^{14}h^{-1}M_\odot$, and $M_\Delta E(z)=M_0(M_{\rm
      gas}E(z))h^{3/2}M_\odot$, respectively. We fix the
    slopes to the self-similar values.
    The X-ray temperature is derived by the volume average of the radial
    temperature profile in the range of $0.2-0.5r_{\rm 500}$
    ($T=T_{\rm 0.2-0.5r_{500}}$; see Zhang et al. 2008). 
    Column (1): Scaling relations.
    Columns (2-4): Normalization of the fit to the relation using weak
    lensing masses. The results for the full sample, undisturbed and
    disturbed clusters are presented. 
    Columns (5-6): Normalization of the mass-observable relations for the 11 clusters and 6 disturbed
    clusters, excluding A1914, respectively. }
\end{table*}

\begin{table*}
  \caption{Intrinsic scatter in mass - observable
    relations}. \label{tab:int_scat} 
  \begin{center}
    \begin{tabular}{l||rrr|rr}
      \hline
      \hline
      Relation  & All 12 clusters
      & 
      & 
      &  w/o A1914
      &  \\
      &   all
      &   undisturbed
      &   disturbed
      &   all 
      &   disturbed  \\
      (1)       &   (2)
      &   (3)
      &   (4)
      &   (5)
      &   (6) \\
      \hline
$M_{500}$--$\Yx$  
& $0.203_{-0.095}^{+0.066}$
& $\le0.283$
& $0.216_{-0.166}^{+0.098}$
& $0.154_{-0.098}^{+0.071}$
& $\le0.225$ \\
$M_{1000}$--$\Yx$  
&$0.203_{-0.078}^{+0.052}$
&$\le0.283$
&$0.241_{-0.166}^{+0.084}$
&$0.173_{-0.081}^{+0.057}$
&$0.243_{-0.204}^{+0.089}$ \\
$M_{2500}$--$\Yx$ 
&$0.245_{-0.088}^{+0.053}$
&$\le0.353$
&$0.306_{-0.191}^{+0.091}$
&$0.245_{-0.097}^{+0.058}$
&$0.329_{-0.280}^{+0.113}$\\
\hline
$M_{500}$--$T_{\rm 0.2-0.5r_{500}}$    
&$0.327_{-0.136}^{+0.087}$
&$\le0.503$
&$0.273_{-0.215}^{+0.123}$
&$0.288_{-0.133}^{+0.093}$
&$\le0.280$\\
$M_{1000}$--$T_{\rm 0.2-0.5r_{500}}$  
&$0.267_{-0.113}^{+0.078}$
&$\le0.364$
&$0.292_{-0.217}^{+0.120}$
&$0.228_{-0.125}^{+0.088}$
&$\le0.318$ \\
$M_{2500}$--$T_{\rm 0.2-0.5r_{500}}$  
&$0.226_{-0.117}^{+0.084}$
&$\le0.295$
&$0.328_{-0.244}^{+0.135}$
&$0.199_{-0.132}^{+0.090}$
&$\le0.411$ \\
\hline
$M_{500}$--$\Mgas$ 
&$0.123_{-0.102}^{+0.068}$
&$\le0.260$
&$\le0.226$
&$0.109_{-0.106}^{+0.062}$
&$\le0.225$ \\
$M_{1000}$--$\Mgas$ 
&$0.160_{-0.096}^{+0.070}$
&$\le0.330$
&$0.207_{-0.188}^{+0.103}$
&$0.147_{-0.104}^{+0.072}$
&$\le0.225$ \\
$M_{2500}$--$\Mgas$ 
&$0.241_{-0.114}^{+0.077}$
&$0.436_{-0.410}^{+0.190}$
&$0.263_{-0.230}^{+0.121}$
&$0.240_{-0.127}^{+0.086}$
&$\le0.248$ \\
\hline
    \end{tabular}
  \end{center}
  \tablecomments{
    Intrinsic scatter, $\sigma_{\ln M}$, in the mass-observable
    relations, using weak-lensing masses, for all 12 clusters and the
    11 clusters excluding A1914 (Sec. \ref{subsec:outlier}),
    respectively. We refer to the mean of the 
    posterior probability distribution of each parameter.
    Column (1): Scaling relations 
    Columns (2-4): Intrinsic scatter for all 12 clusters, undisturbed
    and disturbed clusters, respectively.  
    Columns (5-6):  Intrinsic scatter for the 11 clusters and 6 disturbed
    clusters, excluding A1914, respectively. }
\end{table*}

\begin{table}
  \caption{Multi-dimensional Fitting for covariance of intrinsic
    scatter of gas mass and temperature at $\Delta$=500} \label{tab:Sint}
  \begin{center}
    \begin{tabular}{c|cc}
      \hline
      12 clusters   & Flat Prior & Gaussian Prior\\
      (1)           & (2)   & (3) \\
      \hline
      $\sigma_{g}$  & $0.132_{-0.086}^{+0.066}$
      & $0.121_{-0.061}^{+0.072}$ \\
      $\sigma_{t}$  & $0.213_{-0.080}^{+0.055}$ 
      & $0.206_{-0.051}^{+0.043}$ \\
      $r$           & $\ge0.185$
      & $\ge0.170$ \\
      ${\mathcal P}(|r'|\ge |r|)$  
      & $\le0.565$
      & $\le0.597$\\
     \hline
    \end{tabular}
  \end{center}
  \tablecomments{The model parameters in the covariance of intrinsic
    scatter for gas mass and temperature, derived from the
    multi-dimensional fitting for the full sample.
    We refer to the mean of the posterior distribution of each parameter.
    The lower limit is at a $68.3\%$ confidence level.
    Column (1) : Model parameters.
    Column (2) : Results with the flat prior $p_{\rm prior}=1$. 
    Column (3) : Results with the Gaussian prior with the best-fit
    intrinsic scatter derived from the independent measurement for
    mass-observable relation. 
    ${\mathcal P}(|r'|\ge |r|)$ denotes the probability that the correlation
    coefficient of the two  random variables for twelve pair
    realizations is higher than the observed one.  
  }
\end{table}

\end{document}